# Mutation rules and the evolution of Sparseness and Modularity in Biological Systems

Tamar Friedlander[1], Avraham E. Mayo[1], Tsvi Tlusty[2,3] and Uri Alon[1,*]

[1]Department of Molecular Cell Biology

[2]Department of Physics of Complex Systems

Weizmann Institute of Science

Rehovot 76100, Israel

[3]Simons Center for Systems Biology,

Institute for Advanced Study,

Princeton NJ 08540, USA.

[*] uri.alon@weizmann.ac.il

## Abstract

Biological systems exhibit two structural features on many levels of organization: sparseness, in which only a small fraction of possible interactions between components actually occur; and modularity – the near decomposability of the system into modules with distinct functionality. Recent work suggests that modularity can evolve in a variety of circumstances, including goals that vary in time such that they share the same subgoals (modularly varying goals), or when connections are costly. Here, we studied the origin of modularity and sparseness focusing on the nature of the mutation process, rather than on connection cost or variations in the goal. We use simulations of evolution with different mutation rules. We found that commonly used sum-rule mutations, in which interactions are mutated by adding random numbers, do not lead to modularity or sparseness except for in special situations. In contrast, product-rule mutations in which interactions are mutated by multiplying by random numbers – a better model for the effects of biological mutations – led to sparseness naturally. When the goals of evolution are modular, in the sense that specific groups of inputs affect specific groups of outputs, product-rule mutations also lead to modular structure; sum-rule mutations do not. Product-rule mutations generate sparseness and modularity because they tend to reduce interactions, and to keep small interaction terms small.

**Introduction**

Biological systems show certain structural features on many levels of organization. Two such features are sparseness and modularity (1–10). Sparseness means that most possible interactions between pairs of components are not found. For example, less than 1% of the possible interactions are found in gene regulation networks of bacteria and yeast (11). The second feature, modularity, is the near-decomposability of a system into modules - sets of components with many strong interactions within the set, and few significant interactions with other sets. Each module typically performs a specific biological function. Modularity is found for example in protein structure (functional domains) (12), in regulatory networks (gene modules, network motifs), and in body plans (organs, systems) - for reviews see (2, 8, 10, 13). While modular networks are essentially sparse – sparse networks are not necessarily modular. Even if interactions are few, they could be evenly distributed and therefore not form modules.

Computer simulations of evolution are used to understand the origin of these structural features. The simulations begin with a set of structures, the elements of the structures are mutated, the fitness of each structure is evaluated according to a given goal, and then the structures with the highest fitness are selected. The most commonly used form of mutation in these simulations is the sum-rule mutation: adding a random number to the value of each element. Such simulations typically find optimal structures which satisfy the goal. However, they generally do not yield modular or sparse structures. Even when starting with a modular solution the simulations typically drift towards non-modular solutions, which are usually much more prevalent and are sometimes better at performing given the goal (14). This leaves open the question of how and why sparseness and modularity evolve in biology.

Several studies have addressed this question by employing different approaches. For example, neutral models suggest that duplicating parts of a network can increase its modularity ("duplication-differentiation" model (15)) or similarly that mutation, duplication and genetic drift (16) can lead to modularity. Modularity in metabolic networks was suggested to arise from a neutral growth process (17, 18). On the other hand, other studies suggest that modularity can be selected for, either indirectly or directly. Modularity has been suggested to be beneficial because it provides dynamical stability or robustness to recombination (19), improves the ability to accommodate beneficial foreign DNA (20), breaks developmental constraints (21), evolves due to selection for environmental robustness (22, 23) or because the same network supports multiple expression patterns (24). Horizontal gene transfer, together with selection for novelty can lead to modularity in the polyketide synthase system (25). It was recently suggested by Clune, Mouret and Lipson that network sparseness and modularity can evolve due to selection to minimize connection costs, as is thought to occur for example in neuron networks (26). Kashtan *et al*. (27–29) found that when goals change with time, such that goals are made of the same set of subgoals in different combinations - a situation termed modularly varying goals (MVG) - the system can evolve modular structure. Each module in the evolved structure solves one of the subgoals, and modules are quickly rewired when the goal changes. Modularly varying goals tested in several model systems, with sum-rule mutations used when applicable (14), showed modularity under a range of parameters. Modularly varying goals also speed up evolution relative to unchanging goals (30), a phenomenon evaluated using analytically solvable models (14). Due to the importance of sparseness and modularity in biology, it is of interest to see if additional mechanisms for their evolution exist. In particular, though attention has been given to the goals and cost functions, little attention has been given to the type of mutation rule used.

Here, we address the role of the mutation rule on the evolution of modularity and sparseness. Most studies that use simulations to study evolution employ a simple rule to specify how mutations change the parameters in the structure that is evolved - namely the 'sum rule', in which a parameter is mutated by adding a random number drawn from a specified distribution. Here, we note that this sum rule is usually not a good description of the effect of cumulative genetic mutations on a given biological parameter. Instead, the effects of mutations are better approximated by product-rule processes. For example, the effect of cumulative mutations on an enzyme's activity is found to be multiplicative (31). Similarly, the effect of mutation on binding of proteins to DNA (32, 33) and proteins to proteins (34–36) is thought to be multiplicative to a first approximation, such that the change in affinity caused by several genetic mutations is approximately the product of the effects of each mutation.

One fundamental reason for the use of product rule to describe the effect of genetic mutations is that mutations affect molecular interactions such as hydrogen bonds. This affects the free energy in an approximately additive way, assuming that the different molecular interactions are independent to a first approximation. Since affinity and reaction rate are exponential in free energy, the effects of cumulative genetic mutations on these parameters are approximately multiplicative. Note that in population genetics, there are different meanings to 'additive' and 'multiplicative' mutations (37), and thus we chose the terms 'sum-rule' and 'product-rule' to avoid confusion.

A related feature of mutations is that they more often reduce the absolute strength of the interaction or activity parameter than increase it (38–40). This asymmetry can be captured using product-rule mutations: for example, multiplying by a random number normally distributed $\mathcal{N}(1,\sigma)$ gives equal probability to multiply by 0.5 or 1.5, which tends to reduce the absolute size of the element; in order to revert a 0.5-mutation, one needs to multiply by a 2-mutation, which is less likely to occur.

To study the role of product-rule mutations, we compare evolution of simple and widely used model structures under sum-rule and product-rule mutations in computer evolution simulations. This is of interest because most simulations of evolution use sum-rules for mutations. We found that product-rule mutations lead to evolution of sparseness without compromising fitness. This relates to the study of Burda *et al*. which used a mutation rule that is approximately product-rule (41). In contrast, we found that sum-rule mutations only lead to sparseness under special conditions, such as when the model parameters are constrained to be non-negative. Furthermore, when the goal is modular, we found that product-rule mutations led to modular structures, whereas sum-rule mutations generally do not. Unlike Kashtan *et al.*, (14, 27, 28) here modularity arises from modular goals without need to change goals over time, and when there is no cost for connections. We study the speed and scaling laws of this process. The basic reason that product-rule mutations lead to sparseness and modularity is that they tend to reduce interaction terms and to keep small interaction terms small and thus cause the evolutionary dynamics to approach structures that have optimal fitness with minimal number of interactions. When goals are modular, this effect, in turn, leads to modular structure.

**Results**

**A simple matrix-multiplication model of transcription networks**

To study the effect of the mutation rule on evolved structures, we use a standard evolutionary simulation framework (42, 43). Briefly, the evolutionary simulation starts with a population of $N$ structures, duplicates them, and mutates each structure with some probability according to a mutation rule (the mutation rules described below will be our main focus). Fitness is evaluated for each structure in comparison to a goal. The fittest individuals are selected by a selection criterion, and the process is repeated, until high fitness evolves (Fig. 1A).

We consider, for simplicity, structures described by continuous-valued matrices. These serve as simple models for biological interactions, where the elements of the matrix $A_{ij}$ are the interaction strengths between components $i$ and $j$ in the system. Evolution entails varying the matrix elements to reach defined goals. Linear matrix models have a long history in modeling of biological systems (41, 44–49). Use of a matrix to describe gene expression is a standard approach. Several studies use matrices to reverse-engineer the underlying network (50). Matrix models have also been used to understand developmental gene regulation, as in the pioneering work of Reinitz in Drosophila (51–53); matrix models were recently used by De-Pace *et al.* to relate the strengths of regulation to the level of gene expression across fruit fly species using detailed gene expression measurements (54).

In the field of modularity, matrix models have been extensively used. Matrix models were used in the pioneering work of Lipson *et al.* (55) and also Wagner *et al.* (24). We previously used a matrix model to analytically study a different route to modularity (14). We evolved the matrix $A$ to satisfy the goal $Au = v$, where $u$ and $v$ are vectors. The fitness is the distance to the goal, $F = -\|Au - v\|$ where $\|.\|$ denotes sum of squares of elements (related to Fisher's geometric model (56)).

Often, biological systems have multiple layers (57) where components in one level – e.g. receptors, send signals to components in the next level, e.g. transcription factors. We model this situation using a matrix multiplication model in which we evolve two matrices $A$ and $B$ towards the goal $AB = G$, where $G$ is a specified matrix that represents an evolutionary goal (Fig. 1B). The fitness in this case is $F = -\|AB - G\|$. Note that there is an infinite number of matrix pairs $A$ and $B$ that satisfy a given goal $G$.

As one concrete biological case, which may be kept in mind to guide the reader, the model can be interpreted in the context of a transcription network: if $A$ is the matrix connecting transcription factor (TF) activities to gene expression, the relationship $Au = v$ means that a vector of TF activities $u$ leads to a vector of gene expression $v$. The matrix element $A_{ij}$ thus represents the regulatory strength of gene $i$ by TF $j$. Similarly, if $B$ is a matrix of interactions between external signals $s$ and TF activities, one finds that the TF activities are $u = Bs$. The matrix element $B_{ij}$ represents the effect of signal $j$ on TF $i$. In total, the output gene expression vector that results from an input vector of signals $s$ is $ABs$. The goal $AB = G$ means that for every set of signals $s$, the gene expression at the output of the system is $ABs = Gs$, where $Gs$ is the desired gene expression profile for input signals $s$ (see Fig. 1B).

**Product-rule mutations lead to sparse structures, sum-rule mutations do not**

We compared sum and product mutation rules in evolving the model systems using an evolutionary simulation. The sum-rule is the commonly used addition of a normally distributed random number to a randomly chosen element of the matrices, which represents a mutation in the intensity of a single interaction between network components,

Sum-rule $\qquad A_{ij} \rightarrow A_{ij} + \mathcal{N}(0, \sigma)$ or $B_{ij} \rightarrow B_{ij} + \mathcal{N}(0, \sigma)$.

We also tested product-rules, in which an element of the matrix is multiplied by a random number. We tested

Product-rule $\quad A_{ij} \rightarrow A_{ij} \cdot \mathcal{N}(\mu, \sigma)$ or $B_{ij} \rightarrow B_{ij} \cdot \mathcal{N}(\mu, \sigma)$.

We study the case of $\mu = 1$, and also cases in which $\mu < 1$ and $\mu > 1$. We also tested symmetric multiplication rules where the random number is log-normally distributed with $\mu = 0$ (see Text S1 for details), and thus has equal chance to increase or decrease the absolute strength of the interaction:

Symmetric product-rule $A_{ij} \rightarrow A_{ij} \cdot \mathcal{LN}(0, \sigma)$ or $B_{ij} \rightarrow B_{ij} \cdot \mathcal{LN}(0, \sigma)$.

All cases gave qualitatively similar results, and most of the data below is for multiplying by $\mathcal{N}(1, \sigma)$. We also tested other forms of mutation distributions, including long tailed distributions that describe experimental data on sizes of mutation effects [Gamma distributions (39), see also (40, 58, 59) and references there], and found that the results are insensitive to the type of distribution used (see Supplementary Movies S1, S2). Similarly, we tested the effect of mutation size, that is the parameter $\sigma$, which we varied between 0.01 and 3, and found that the results are insensitive to this parameter. The evolutionary simulation and parameters are described in the Methods Section below.

To demonstrate the effect of the mutation rule, we begin with a very simple model, namely a structure with two elements, $x$ and $y$, with fitness $F = -(x+y-1)^2$. The optimal solutions lie on a line in the $(x, y)$ plane, namely $x + y = 1$ (Fig. 2A). Evolutionary simulations reach this line regardless of the mutation rule. Populations under sum-rule mutations evolve and spread out over the line. In contrast, product-rule mutations lead to solutions near the axes, either $(x=0, y=1)$, or $(x=1, y=0)$. In other words, they lead to solutions in which one of the elements is close to zero – these are the sparsest solutions that satisfy the goal (see Fig. 2A, Supplementary Movies S1, S2 and Figs. S10-S11).

The intuitive reason for the sparseness achieved by product-rule mutations is that once they are near a zero element, the size of the next mutation will be small (since it is a product of the element with a random number). Thus, the effective diffusion rate decreases (see Text S1). Strictly zero terms are fixed-points and near-zero terms remain small under mutations - so that the population becomes concentrated near zero elements. Sum-rule mutations, in contrast, show a constant drift rate regardless of the value of the elements. A full analytical solution of the dynamics of this simple model can be obtained by means of Fokker-Planck equations (see Text S1, Section 1), in excellent agreement with the simulations.

We tested product-rule mutations also in the matrix-multiplication model, using as goals full rank matrices $G$. In numerical simulations, we refer to terms that are relatively small (<0.1% of the average element in $G$) as "zero terms", because strictly zero terms are not reached in finite time. We find that product-rule mutations lead to sparseness: matrices $A$ and $B$ with the highest number of zeros possible while still satisfying the goal. In contrast, sum-rule mutations result in non-sparse solutions $A$ and $B$ with non-zero elements (Fig. 3).

The sparse solutions found with product-rule mutations have many zero terms, whose number can be computed by means of the LU decomposition theorem of linear algebra. The LU decomposition expresses a nonsingular matrix as a product of an upper triangular matrix and a lower triangular matrix (60)). The total number of zeros in $A$ and $B$ is the number of zero elements in the LU decomposition of $G$. This number can be calculated exactly: for a given full rank matrix $G$ of dimension $D$ with no zero elements, the maximal number of zeros in $A$ and $B$ together is $D^2 - D$ (for proof see Text S1). This result is found in our simulations.

The zeros are distributed between $A$ and $B$ in various ways in the different simulations: Sometimes $A$ and $B$ are both (upper and lower) triangular, each with $(D^2 - D)/2$ zero elements. Other runs show one full matrix with no zeros and the other a diagonal matrix with $D^2 - D$ zeros. All other distributions of zeros are also found (Fig. 2B-C; Fig. S16 for comparison with sum-mutations). When $G$ is full rank and has $k$ zeros, the total number of zeros in the evolved matrices $A$ and $B$ is $D^2 - D + k$, again the maximal possible number of zeros in matrices that show optimal fitness (for proof see Text S1).

We note that there is a special situation in which sum-rule mutations can also lead to sparseness in the present models. This occurs when the models are constrained to have only non-negative terms $A_{ij}, B_{ij} \geq 0$. In this case, the sum rule, constrained to keep terms non-negative – for example, by using $A_{ij} \to \max(0, A_{ij} + \mathcal{N}(0, \sigma))$, can also lead to sparseness. This relates to known results from non-negative matrix factorization (61). However, in general biological models, structural terms are expected to be both negative and positive, representing, for example, inhibition and activation interactions between components. Our mechanism for the evolution of sparseness and modularity is different from non-negative matrix factorization and works regardless of the sign of the interaction terms (see for examples Fig. S15 and Supplementary Movies S1, S2).

**When the goal is modular, product-rule mutations lead to modular structure; sum-rule mutations do not**

Up to now, we considered goals $G$ which are described by general matrices. We next limit ourselves to the case where the goals $G$ are described by matrices which are modular, for example, diagonal or block diagonal matrices. The main result is that when the goals are modular, the evolved structures $A$ and $B$ are also modular if mutations are product-rule; in contrast, sum-rule mutations lead to $A$ and $B$ that are not modular despite the fact that the goal is modular.

We first define modular structures and modular goals in the context of the present study. Modular structures are structures that can be decomposed into sets of components, where each set shows strong interactions within the set and weak interactions with other sets (1, 2, 10, 62) (Fig. 1B). Here, modular structure means block-diagonal matrices. For ease of presentation, we first consider the most modular of structures – namely diagonal matrices. We define modularity by $M = 1 - \langle |n| \rangle / \langle |d| \rangle$ where $\langle |n| \rangle$ and $\langle |d| \rangle$ are the mean absolute value of the non-diagonal and diagonal terms respectively, and where we permute rows/columns to maximize modularity $M$ (same permutation for rows of $A$ and columns of $B$, see Text S1). Thus, a diagonal matrix has $M = 1$, and a matrix with diagonal and non-diagonal terms of similar size has $M$ close to zero.

Modular goals are goals which can be satisfied by a modular structure. Modular goals in the present models are represented by diagonal or block-diagonal goal matrices $G$. These goals, in the biological interpretation of transcription networks (Fig. 1B), are goals in which each small set of signals affects a distinct set of genes, and not the rest of the genes. For example, the signal lactose affects the *lac* genes in *E. coli*, whereas a DNA damage signal affects the *SOS* DNA-repair genes, with little crosstalk between these sets. Other examples for biological goals that are modular are sugar metabolism (63) and the tasks of chemotaxis and organism development (see detailed discussion in (27)). All are composed of several sub-tasks that are associated with different sets of genes.

We note that a modular goal does not necessarily lead to modular structures. For example the goal $G = I$ is modular since $I$ is the diagonal identity matrix. This modular goal can be satisfied by a product of infinitely many pairs of non-modular matrices $AB$. In fact, for every invertible $A$, the inverse $B = A^{-1}$ satisfies the goal. As a result, the vast majority of the possible solutions are non-modular (modular solutions have measure zero among possible solutions to $AB = G$). In line with this observation, we find that simulations with sum-rule mutations lead to solutions with optimal fitness ($AB = G$), but with non-modular structure $A$ and $B$ (Fig. 3, Fig. S16).

In contrast, we find that product-rule mutations lead to modular structures $A$ and $B$, for a wide range of parameters. For the goal $G = I$, the evolved $A$ and $B$ are both diagonal matrices, with elements on the diagonal of $A$ that are the inverse of the corresponding elements on the diagonal of $B$. Thus $AB = G$. Similar results are found if the goal is nearly modular (e.g. diagonal with small but nonzero off-diagonal terms): in this case, the evolved $A$ and $B$ are both nearly diagonal (Fig. S14).

We also studied block-modular goals. In this case, product-rule mutations led to block-modular matrices $A$ and $B$, with the same block structure as the goal matrix $G$ (Fig. 2C). Each of the blocks in the matrices $A$ and $B$ had the maximal number of zeros possible so that the product of the two blocks is equal to the corresponding block in the goal matrix $G$ (the total number of zeros is equal to that in the LU decomposition of each block) – compared to Fig. S16 (block-diagonal goal with sum-rule mutations).

It is important to note that in order to observe the evolution of modularity in the present setting, the selection criteria should not be too strict, otherwise non-modular solutions cannot be escaped effectively (Text S1). In other words, overly strict selection does not allow the search in parameter space needed for product-rule mutations to reach near-zero elements. In the present simulations, we find evolution of modularity using standard selection methods including tournament, elite (truncation) and continuous Boltzmann-like selection (see Methods, and Text S1 for analysis of sensitivity to parameters).

**Time to evolve modular structure increases polynomially with matrix dimension**

We studied the dynamics of the evolutionary process in our simulations with product-rule mutations. We found that over time, fitness and modularity both generally increase, until a solution with optimal fitness and maximal modularity is achieved. We found that the matrix multiplication model often shows plateaus where fitness is nearly constant, followed by a series of events in which fitness improves sharply (Fig. 4) (22, 64). In these events, modularity often drops momentarily. Analysis showed that the plateaus represent non-modular and sub-optimal structures. A mutation occurs which reduces modularity but allows the system to readjust towards higher fitness, and then to regenerate modularity.

We also tested the time to reach high fitness solutions, and its dependence on the dimension of the matrices $D$. The time to high fitness solutions depends on the settings of the simulations: initial conditions, selection criteria and mutation rates and size, and the stopping criteria of the simulations. Here we present results in which time to high fitness was measured as the median time over repeat simulations to reach within 0.01 of optimal fitness, with product mutation rule $\cdot \mathcal{N}(1,0.1)$ and probability of mutation per element that is dimension-independent ($p = 5 \times 10^{-4}$). Initial conditions were matrices with small random elements ($\mathcal{U}(0,0.05)$). The time to high fitness increased approximately as $D^{\theta_1}$ with $\theta_1 = 1.40$+/-0.01 and the time to modularity (see Methods for definition) increased as $D^{\theta_2}$ with $\theta_2 = 1.21$+/-0.04 (Fig. 5).

**Discussion**

We found that product-rule mutations lead evolution towards structures with the minimal number of interaction terms that still satisfy the fitness objective. Thus, product-rule mutations lead to sparseness. When the goal is modular, product-rule mutations lead to modular structure. This is in contrast to sum-rule mutations, which lead, under the same conditions, to non-sparse and non-modular solutions.

The mechanism by which product-rule mutations lead to sparseness and modularity is that near-zero interaction terms are kept small by product-rule (but not sum-rule) mutations. A second effect is mutation asymmetry, where it is more likely to reduce an interaction than increase it. However, using a symmetric product-rule (multiplying by a number drawn from a symmetric log-normal distribution) combined with selection still leads to sparseness and modularity, because selection also breaks the symmetry. Once a parameter becomes small, product rule mutations keep it small (as opposed to sum-rule mutations). This creates a dynamic 'trap' in which the steady state distribution of phenotypes is highly enriched with near zero parameters. Thus, the mutational asymmetry effect is not essential for the present conclusions (see Figs. S10-S11 and Section 2 in Text S1). Furthermore, in special situations a sum-rule can also lead to sparseness, namely if the structural terms in the model are constrained to be non-negative. We note that sparseness can also be enhanced in some networks due to physical constraints such as spatial/geometric limitations in networks that describe protein structure (12) or neuron wiring networks (65).

We used a simple but general model of biological systems, namely linear matrix models, and matrix multiplication models. These models have been widely used to describe gene regulation, neuronal networks, signal transduction and other systems (41, 44–49, 66, 67). The matrix multiplication model is a commonly used model for three layer systems, such as signals → transcription factors → genes. As in many biological models, many combinations of parameters can achieve the same goal.

We believe that the present mechanism has generality beyond the particular model used here. Consider a general map $H$ between a coarse-grained genotype $G$ (described as a set of biochemical parameters and interaction parameters) and phenotype $P$, $P = H(G)$. The optimal phenotype $P^*$ is obtained by a manifold of different genotypes $G^*$. Given reasonably strong selection relative to genetic drift and mutation, evolutionary dynamics will reach close to this manifold. One can then ask how the mutation rule affects evolutionary dynamics along this manifold. Sum-rule mutations lead to a random walk on the manifold that does not prefer regions with small parameters, whereas product-rule mutations lead to solutions with the maximal number of zero (very small) parameters: once evolution comes close to a zero parameter, product-rule mutations keep that parameter small.

Product-rule is a more realistic description of the effect of cumulative genetic mutations on a biochemical parameter than sum-rule mutations, because of the nature of biological interactions. The effect of genetic mutations was also shown in several experimental studies to be asymmetric (for example (38–40)), with bias to decrease interactions, enzymatic activity (38) or body size (40). The discussion of symmetric product rule mutations (that is - multiplying by log-normally distributed random numbers) is given here for completeness, and not because of biological relevance. Further studies can use other microscopic models for mutations (such as Ising-like models for bonds between macromolecules (41, 68)), and explore the effect of mutations that set interactions to near-zero with large probability. Due to the inherent product-rule nature of biological mutations, we could not think of experimental tests that can compare sum-rule to product-rule mutations, beyond computer simulations or experiments in the realm of electronics (69, 70) or mechanics (71).

The present mechanism does not exclude previous mechanisms for the evolution of modularity. In fact, it can work together with other mechanisms and enhance them. For example, in Kashtan *et al*. (14, 27, 29, 30), modularity evolved when the modular goal changed over time (MVG mechanism). In the present study, no change of the goal over time is required. Using product-rule mutations in the models of Kashtan *et al* (instead of the original sum-rule mutations) is expected to enhance the range of parameters over which modularity evolves. Supporting evidence was recently provided by Clune *et al*. (26) that demonstrated how a different mechanism for the evolution of sparseness significantly enlarges the range of parameters over which the MVG mechanism produces modular structures. Another difference from some previous studies is that modularity evolves here with no need for an explicit cost for interaction terms in the fitness function (14, 27, 29, 30, 72). Adding such a cost, as in Clune *et al*., (26) would likely enhance the evolution of sparseness and modularity. It would be intriguing to search for additional classes of mechanisms to understand the evolution of sparseness, modularity, and other generic features of biological organization (73).

**Materials and Methods:**

**Evolutionary simulation**

Simulation was written in Matlab (all source codes, data and analysis scripts are freely available in a permanent online archive at XXXXX (will be deposited upon manuscript acceptance)) using standard framework (42, 43). We initialized the population of matrix pairs by drawing their $N \cdot 2D^2$ terms from a uniform distribution. Population size was set as $N = 500$. In each generation the population was duplicated. One of the copies was kept unchanged, and elements of the other copy had a probability $p$ to be mutated – as we explain below. Fitness of all $2N$ individuals was evaluated by $F = -\|AB - G\|$, where $\|\cdot\|$ denotes the sum of squares of elements (56). The best possible fitness is zero, achieved if $AB = G$ exactly. Otherwise, fitness values are negative. In the figures we show the absolute value of mean population fitness, which is the distance from maximal fitness (Fig. 3-4, S12), or the normalized fitness $|F|/(|G|+|AB|)$ (Fig. 5). The goal matrix was either diagonal $G = 2 \times I$, nearly-diagonal (diagonal matrix with small non-diagonal terms), block-diagonal or full rank with no zero elements. $N$ individuals are selected out of the $2N$ population of original and mutated ones, based on their fitness (see below). This mutation–selection process was repeated until the simulation stopping condition was satisfied (usually when mean population fitness was within 0.01 of the optimum).

**Mutation**: We mutated individual elements in the matrix. We set mutation rate such that on average 10% of the population members were mutated at each generation, so the element-wise mutation rate was $\sim \frac{0.1}{2D^2}$. This relatively low mutation rate enables beneficial mutants to reproduce on average at least 10 generations before they are mutated again. In simulations where we compared dependence on matrix dimension (Fig. 5) we used the same mutation rate at all dimensions, generally the one that pertains to the highest dimension used in the simulation.

We randomly picked the matrix elements (in both $A$ and $B$) to be mutated. Mutation values were drawn from a Gaussian distribution (unless otherwise stated). For sum-rule mutations, this random number was added to the mutated matrix value: $A_{ij} \to A_{ij} + \mathcal{N}(0, \sigma)$ or $B_{ij} \to B_{ij} + \mathcal{N}(0, \sigma)$, and for product-rule mutation, the mutated matrix element was multiplied by the random number: $A_{ij} \to A_{ij} \cdot \mathcal{N}(\mu, \sigma)$ or $B_{ij} \to B_{ij} \cdot \mathcal{N}(\mu, \sigma)$. Mean mutation value $\mu$ was usually taken as 1, however we also tested other values of $\mu$ (both larger and smaller than 1) and other mutation distributions (Gamma and log-normal) and results remained qualitatively similar, although the time-scales changed. In most simulations shown here we used $\sigma = 0.1$ (unless stated otherwise). Fitness convergence and its time scale depend on the mutation frequency and size, as demonstrated in our sensitivity test (Text S1).

**Selection methods**: We tested 3 different selection methods and all gave qualitatively very similar results with only a difference in time scales. Most results presented here were obtained with tournament selection with group size S=4 (see (43) chap. 9). We also tested truncation-selection (elite) (42) and proportionate reproduction with Boltzmann-like scaling (41, 55, 74). For a detailed description see Text S1.

**Definition of modularity**: if the goal is diagonal, we define modularity as $M = 1 - \langle |n| \rangle / \langle |d| \rangle$ where $\langle |n| \rangle$ and $\langle |d| \rangle$ are the mean absolute value of the non-diagonal and diagonal terms respectively. At each generation, the $D$ largest elements of each matrix (both $A$ and $B$), were considered as the diagonal $\langle |d| \rangle$ and the rest $D^2 - D$ terms as the non-diagonal ones $\langle |n| \rangle$. Averages were taken over matrix elements and over the population. This technique copes with the unknown location of the dominant terms in the matrices, which could form any permutation of a diagonal matrix. Thus, $0 \leq M \leq 1$: a diagonal matrix has $M = 1$, and a matrix whose terms are all equal has $M = 0$. Since we choose the largest elements to form the diagonal, negative values of $M$ do not occur. When the goal is non-diagonal, one can use standard measures for modularity such as (49) [not used in the present study].

**Calculation of time to modular structure**: To estimate the time when modular structure is first obtained, we used the following approximation for fitness value with diagonal goal. Assume that $A$ and $B$ are $D$-dimensional matrices consisting of 2 types of terms: diagonal terms all with size $d$ and non-diagonal terms all with size $n$ and that the goal is $G = g \times I_{D \times D}$. The fitness then equals:

$$-F = D[d^2 + (D-1)n^2 - g]^2 + D(D-1)(2dn + (D-2)n^2)^2.$$

We collect terms by powers of $n$, and obtain a constant term and terms with powers $n^{2,3,4}$. Modular structure is obtained when the solution has the correct number of dominant terms at the right location and their size is approximately $d^2 \cong g$. At the beginning of the temporal trajectory, when non-diagonal elements are relatively large, $F$ is dominated by the $O(n^4)$ term. When a modular structure emerges, non-diagonal elements become relatively small, and the dominant term remaining in $F$ is $O(n^2)$. Our criterion for determining time to modular structure was the time when the $O(n^2)$ term first became dominant, i.e. when $F - n^2(\cdots) < n^2(\cdots)$.

**Acknowledgements**

We thank Yuval Hart, Adam Lampert, Omer Ramote, Hila Sheftel, Oren Shoval and Pablo Szekely for critical reading of the manuscript; and Amos Tanay, Dan Tawfik, Nadav Shnerb and Gheorghe Craciun for useful discussions. We thank the two anonymous referees for helpful comments and suggestions that improved our manuscript. The research leading to these results received funding from the Israel Science Foundation and the European Research Council under the European Union's Seventh Framework Programme (FP7/2007-2013) /ERC Grant agreement n° 249919. U. A. is the incumbent of the Abisch-Frenkel Professorial Chair. T.F. acknowledges a Clore fellowship.

**Figure Captions**

**Fig. 1 – Evolutionary simulation scheme, and definition of model**. **(A)** Simulation was initiated by randomly choosing $N$ population members each consisting of 2 matrices $A$ and $B$. The next steps were repeated at each generation until the stopping condition was satisfied: the population was duplicated, one copy was kept unchanged and the other was mutated with probability $p$. Mutation could be either sum-rule or product-rule. Fitness of all $2N$ members (original and mutated) was evaluated by the distance of the product $AB$ from a desired goal matrix $G$, $F = -\|AB - G\|$, where $\|\cdot\|$ denotes the sum of squares of terms which is the square of $L_2$ (Frobenius) norm. $N$ individuals were selected according to their fitness. Several selection methods were employed (see Text S1 for details). The simulation was stopped when the mean population fitness reached a value which was within a preset difference from the optimal fitness (usually 0.01). **(B)** Model represents a three layer network with a linear transformation function. Input signals $s$ are transformed to intermediate layer activities (transcription factors) $u$ with $u = Bs$. The intermediate layer is then transformed to output layer (gene expression) $v = Au = ABs$. Modularity means block or diagonal structure of the matrices, corresponding to signals that affect only subsets of intermediate and output nodes.

**Fig. 2 – Product-rule mutations reach sparse solutions, sum-rule ones do not. (A)** We demonstrate the difference between sum-rule and product-rule mutations in a simple 2–variable system $(x, y)$, where the goal is that $x + y = 1$. The optimal solutions lie on the line $y = 1 - x$. We compare solutions to this problem achieved by 3 different mutational schemes. Sum-rule mutations ($+\mathcal{N}(0, 0.5)$ red circles) provide solutions that are spread along the line. In contrast, solutions achieved with both Gaussian product-rule ($\times \mathcal{N}(1, 0.5)$ blue diamonds) and log-normal product-rule ($\times \mathcal{LN}(1, 0.5)$ green squares) are concentrated near the intersection with the axes, i.e. near either (0,1) or (1,0). Since one coordinate is near zero, these are sparse solutions. Inset illustrates the solutions obtained with Gaussian product-rule mutations, demonstrating that matrix values can be negative as well as positive. Evolutionary simulation parameters were $p = 0.5$, $N = 500$, selection scheme was Boltzmann-like selection with $\beta = 10$. Simulations initiated utilizing random matrices with elements $\mathcal{U}(0, 0.05)$. **(B)** Sparse solutions evolve in the matrix-multiplication model under product-rule mutations in response to a full (non-zero) goal matrix $G$. The solutions have the maximal number of zeros while still satisfying the goal. Zeros are distributed between the two matrices $A$ and $B$. Shown are the possible configurations of $A$ and $B$ for matrices of dimension $D = 3$, in which 6 zeros are distributed between the two matrices $A$ and $B$. **(C)** In general, if the goal is block diagonal, each of its blocks can be decomposed separately into blocks of $A$ and $B$, such that each block has the maximal number of zeros possible. Here we show an example in $D = 4$, where $G$ has 2 blocks of 2X2. The evolved $A$ and $B$ are such that each of their blocks is either an upper or a lower triangular matrix. Color represents numerical value (white = zero).

**Fig. 3 – Product-rule mutations lead to modular structure under modular goal, sum-rule mutations do not.** **(A)** Both sum-rule and product-rule mutations reach high fitness towards the goal $G = 2I$. **(B)** Product-rule mutations reach high modularity, but sum-rule mutations do not.

Simulations are in the matrix-multiplication model, matrix dimension $D=4$. Examples of matrices drawn from the simulations are shown, with gray scale corresponding to element absolute value (white=zero). Fitness reaches a value of 0.01 due to constantly occurring mutations. Evolutionary simulation parameters are: sum-rule mutation size $\mathcal{N}(0,0.05)$, product-rule mutation size $\mathcal{N}(1,0.27)$, $p = 0.0031$, tournament selection $s=4$.

**Fig. 4 - Escape from a fitness plateau entails a temporary decrease in modularity.** Mean distance from maximal fitness as a function of time in the matrix-multiplication model with product-rule mutations, towards a diagonal goal. Note the plateau in the dynamics. Matrices and their modularity drawn from the simulations at different time-points (designated by black points) are shown, with gray scale corresponding to element absolute value (white=zero). Inset: mean modularity of population (red curve), showing a sharp decrease at the time of escape from the plateau (same time points are shown). In order to escape the plateau ("break point"), the circled terms in $A$ and $B$ are changed. This occurs through a simultaneous increase of the new term and decrease of the old one, such that temporarily modularity is decreased (see inset). Finally, the correct arrangement of terms is attained ("escape") and modularity increases again. Fitness reaches a value of 0.01 due to constantly occurring mutations.

**Fig 5: Time to high fitness and modularity grows polynomially with dimension. (A)** Normalized distance to maximal fitness $|F|/(|G|+|AB|)$ as a function of generations in the matrix-multiplication model evolved towards $G=2I$, for matrix dimensions $D=3$ to $10$. Each color represents a different value of $D$. Curves typically have $D$ "steps", where each step corresponds to the build-up of an additional significant term. **(B)** Modularity in the same simulations. **(C)** Median time (generations) to high fitness (distance from maximal fitness < 0.01) as a function of the dimensions of the matrices $D$ goes as $T \sim D^{\theta_1}$ with $\theta_1 = 1.41$ [1.40, 1.42] [CI 5%, 95%]. **(D)** Median time (generations) to modular structure (see methods for dimension dependent criterion for high modularity) goes as $D^{\theta_2}$ with $\theta_2 = 1.20$ [1.16, 1.23]. Initial conditions are random matrices with small elements drawn from $\mathcal{U}(0,0.1)$. Element-wise mutation rate $p$ at all simulations was 0.0005; product-rule mutations normally distributed $\mathcal{N}(1,0.1)$. See Text S1 for details on error calculation in **C**-**D**.

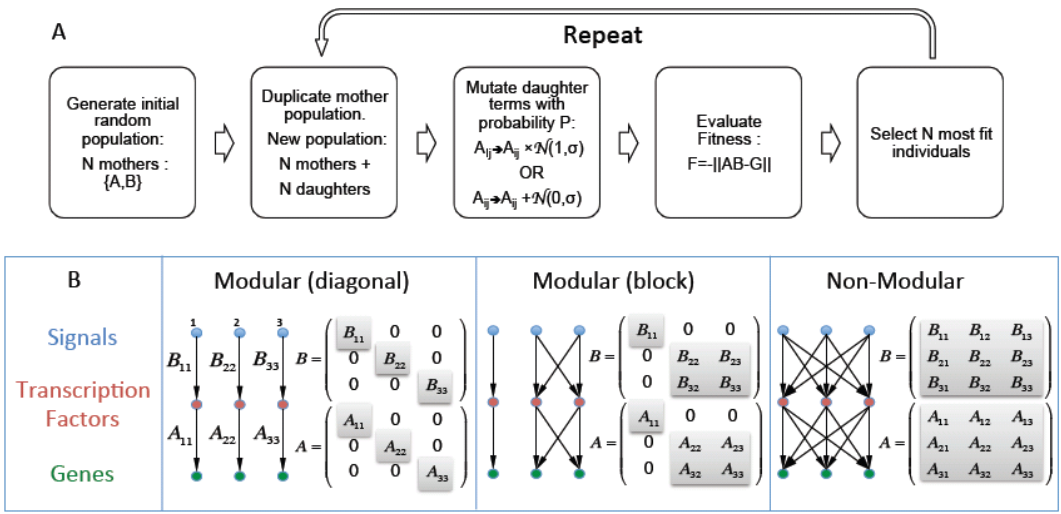

**Fig. 1**

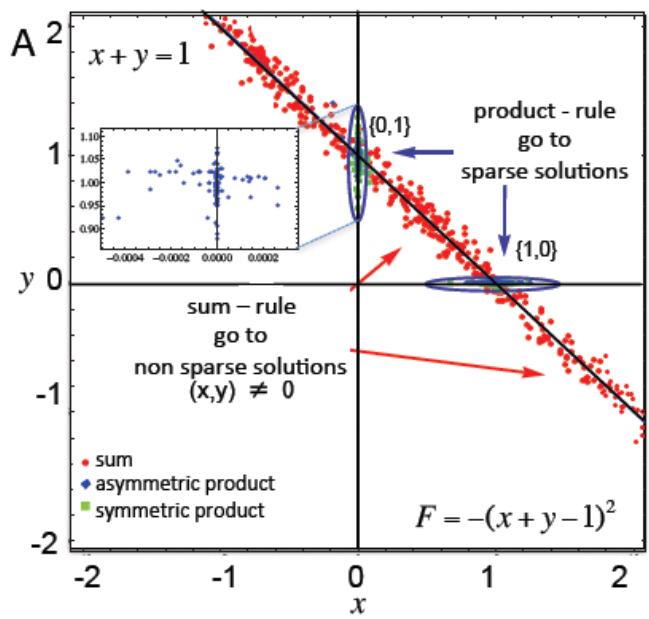

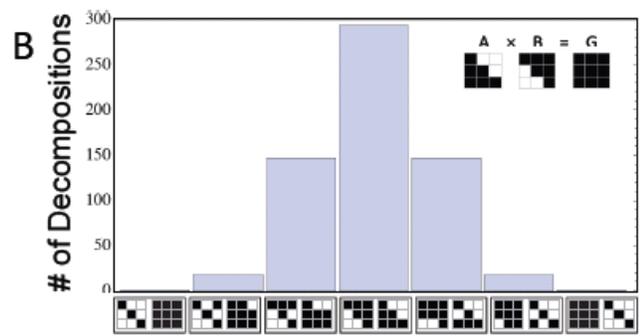

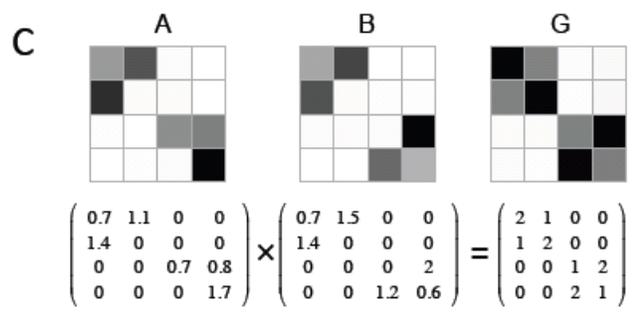

**Fig. 2**

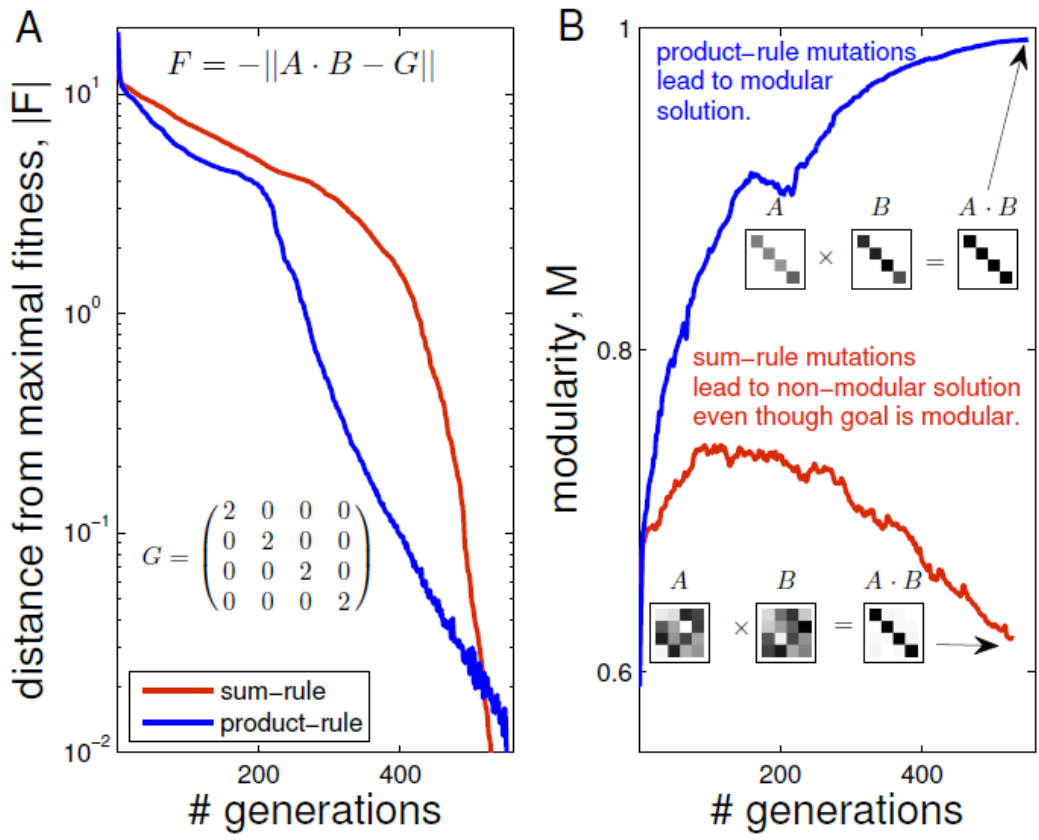

**Fig. 3**

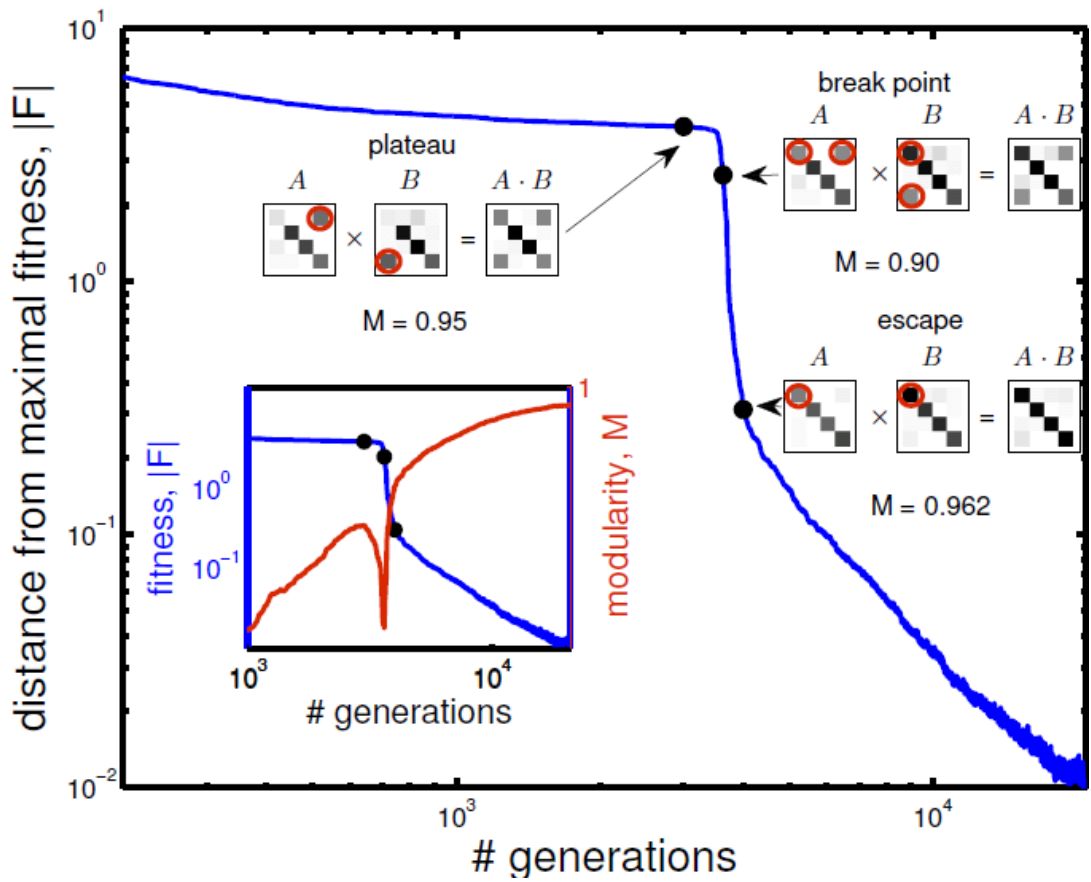

**Fig. 4**

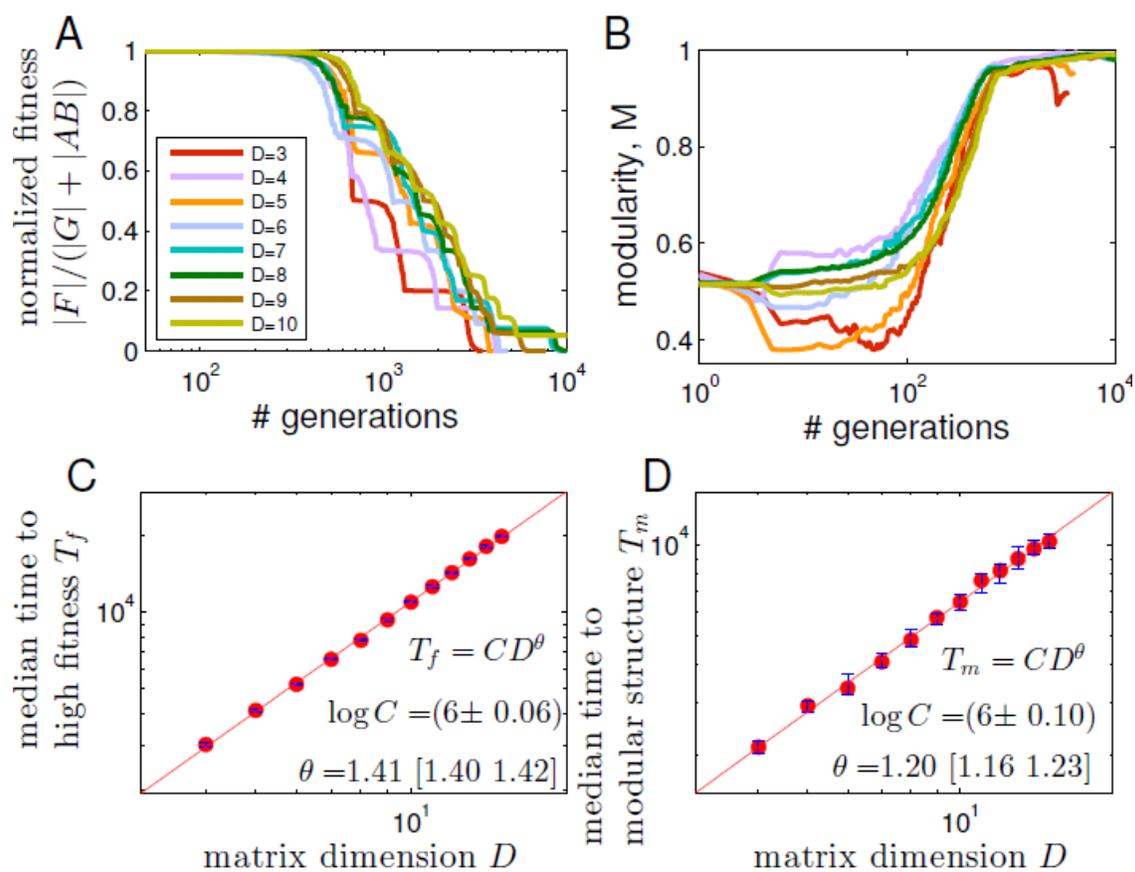

**Fig. 5**

# Mutation Rules and the Evolution of Sparseness and Modularity in Biological Systems

Tamar Friedlander, Avraham E. Mayo, Tsvi Tlusty and Uri Alon

**Supporting Information**

# Table of Contents

1. **Analytical solution and simulations of toy model**

To gain better insight into the effect of the product mutational mechanism we studied a simple toy model. We showed that the effect of sum-mutations is equivalent to free diffusion in isotropic medium along equal fitness lines with no preference to any specific solution on this line. The effect of product-mutations in contrast is described by diffusion in the log-transformed parameter domain. In the original domain the population is log-normally distributed and asymptotically approaches zero. If selection were absent this mutational mechanism would nullify all network interactions. However the combination of the product-mutations with selection for achieving a certain goal results in solutions with maximal number of zeros that still satisfy the goal. The dynamics and type of solutions demonstrated in this model is representative of those we obtained in simulations of the more complex matrix-multiplication model described in the main text.

We study the simplest model in which there is an excess degree of freedom, namely a two variable model such that a modular solution is enabled. We assume that the fitness function depends on the two variables only through their sum. That is, the population exists in a 2-variable space $(x, y)$ and its goal is to reach the line where $x + y = 1$. All points on this line are equally fit, but only two of them - the intersections with the axes (0,1) and (1,0) are sparse. This is because we interpret the variables as interaction intensities between network components and a sparse network is one in which some interactions are zero. Fitness is evaluated by the square distance from this line $F(x, y) = -(x + y - 1)^2$. Although the model does not include terms which depend on products of variables (as in the more general model that we simulated) it is still useful for comparing the effects of the sum and product mutational schemes on the evolutionary dynamics.

In our analysis of the toy model we made a number of simplifying assumptions with respect to the simulations. First we assume continuous time, instead of the discrete generations in the evolutionary simulation. We also take the limit of infinitely large mutation rate with infinitesimally small mutation size, such that their product is finite, and can be described by a diffusion coefficient (compare to (1)). Furthermore we take the limit in which population size is very large, so that fluctuations and random drift due to finite size effect are negligible.

The population dynamics is naturally decomposed into two axes: along equal-fitness lines dynamics is mutation (diffusion) dominated; in perpendicular to such lines it is determined by a combination of mutation and selection (see Fig. S1). We solve analytically the dynamics of the diffusion-dominated axis, and quantify the speed with which sparse solutions are approached with product-mutations. We also obtain a steady-state solution for the mutation-selection axis, showing that it obeys a Boltzmann distribution. We demonstrate our findings by detailed stochastic simulations (see below), showing good agreement with the analytical solutions.

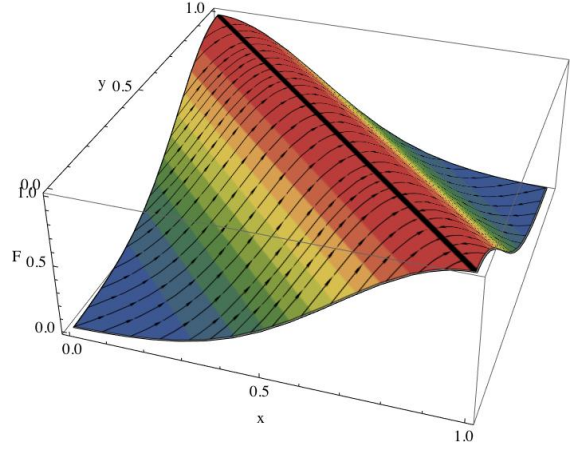

**Fig S1: Decomposition of the x, y problem into its natural axes.** Color represents fitness. Along equal fitness lines only mutation (diffusion) plays a role. Arrows show the other axis along which both selection and mutation are active. The maximal fitness line $x + y = 1$ is shown in bold.

### Dynamics under sum-rule mutation

The mutation-selection dynamics of the population is approximately captured by the Fokker-Planck (abbreviated below 'FP') equation (2). We denote by $P(x, y, t)$ the population distribution at time t. It is subject to the potential $F(x, y) = -(x + y - 1)^2$ (selection), and diffusion with coefficient D (mutation):

$$\partial_t P(x, y, t) = D \nabla^2 P(x, y, t) - \nabla \cdot \big( P(x, y, t) \nabla F(x, y) \big).$$

The Fokker-Planck equation is the continuous second order approximation of the more general master equation describing the dynamics of a population subject to probabilistic transitions between states. For example in the 1-dimensional case the master equation takes the form:

$$P(x, t + \Delta t) = P(x, t) + \sum_z w(z, x) P(z, t) - \sum_z w(x, z) P(x, t)$$

where $w(z, x)$ is the transition probability from $z$ to $x$. By second order approximation we neglect transitions between grid points which are far from one another (i.e., mutations generally have a small effect). This assumption translates to $w$ being a narrow function of its arguments and is common practice in the literature (see for example (3)). We also assume that the transition probability $w(z, x)$ depends only on the difference in fitness between $x$ and $z$. For example $w(z, x) = e^{(F(z) - F(x))}$. With this we obtain:

$$P(x, t + \Delta t) = P(x, t) + e^{(F(x+\Delta x) - F(x))} P(x + \Delta x, t) + e^{(F(x-\Delta x) - F(x))} P(x - \Delta x, t) - (e^{(F(x) - F(x+\Delta x))} + e^{(F(x) - F(x-\Delta x))}) P(x, t).$$

Expanding this equation to first order in $\Delta t$ and second order in $\Delta x$, and rearranging we obtain the Fokker-Planck type equation as above with the diffusion coefficient D given by $\Delta x^2/\Delta t$ as usual (3).

Taking into account the specific form of the fitness function F in this problem, it is convenient to make the following coordinate transformation:

$$u = x + y$$

$$v = x - y$$

With that, the potential and the FP equation transform to:

$$\tilde{F}(u) = -(u-1)^2$$

$$\partial_t \tilde{P}(u,v,t) = 2D\,\partial_u\,\partial_u\tilde{P} + 2D\,\partial_v\,\partial_v\tilde{P} - \partial_u\tilde{F}\,\partial_u\tilde{P} - \tilde{P}\,\partial_u\,\partial_u\tilde{F}.$$

This equation can be solved by separation of variables. We assume that the population already converged to the line of optimal fitness, $u = 1$. Therefore the time dependence of P enters only through $\tilde{P}(u,v,t) = V(v,t)U(u)$. The equation then reads:

$$\frac{\partial_t V - 2D\,\partial_v\,\partial_v V}{V} = \frac{2D\,\partial_u\,\partial_u U - \partial_u\tilde{F}\,\partial_u U - U\,\partial_u\,\partial_u\tilde{F}}{U} = \alpha = \text{const}.$$

For $\alpha \neq 0$ the equation in V describes a population growing at rate $\alpha$. In our simulations we keep the population size constant, thus we set here $\alpha = 0$, and obtain the following equations for $U(u)$ and $V(v,t)$:

$$\partial_t V - 2D\,\partial_v\,\partial_v V = 0$$

$$2D\,\partial_u\,\partial_u U - \partial_u\tilde{F}\,\partial_u U - U\,\partial_u\,\partial_u\tilde{F} = 0$$

Thus, the dynamics of the V component is described by the diffusion equation with diffusion coefficient $2 \cdot D$. Its solution is the normal distribution with variance that grows linearly in time (2, 4):

$$V(v,t) = \frac{1}{\sqrt{8\pi Dt}}\,e^{-\frac{v^2}{8Dt}}.$$

The solution for the U component is the Boltzmann distribution with potential F and effective 'temperature' D (2):

$$U(u) = \frac{e^{F(u)/D}}{\int e^{F(u)/D}}.$$

This steady state solution manifests the balance between selection and mutation by the ratio F/D. A low (high) F/D ratio results in a wide (narrow) distribution around the line of maximal fitness. In summary, along characteristics perpendicular to the optimal fitness line, the population density decays with the distance from the maximal fitness line; along characteristics parallel to this line the population diffuses freely. Our conclusions apply to any potential of the form $\tilde{F}(u) = g[(u-1)^2]$. In the simulations, we used a specific function (see below).

This behavior is demonstrated in Figs. S2-S4, showing the distributions of $x + y$ and $x - y$ and the time-dependence of their moments obtained in simulations with sum-mutations.

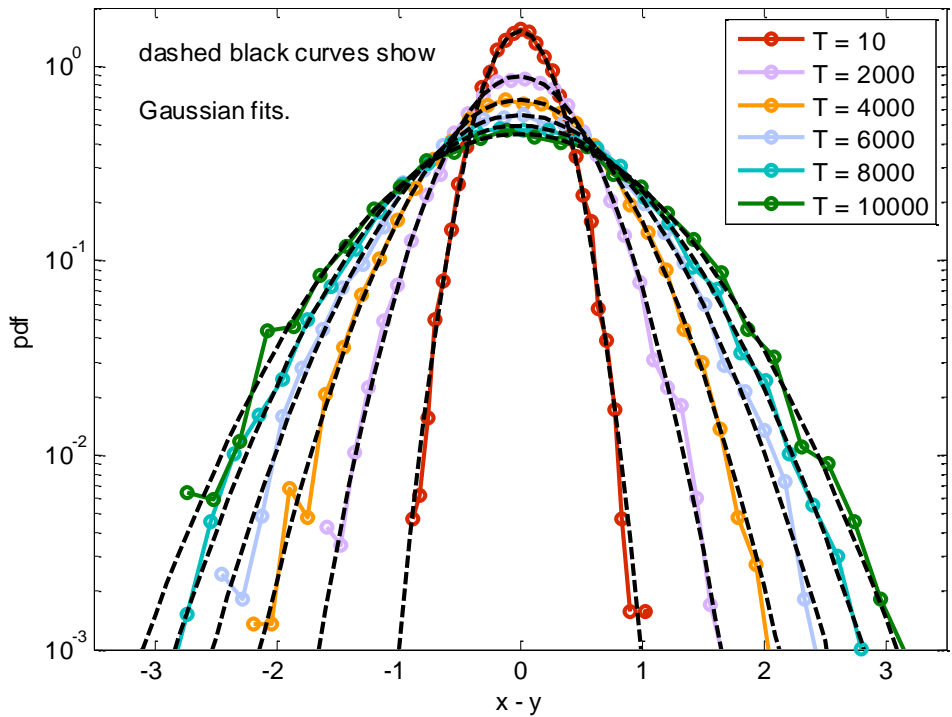

**Fig S2: $x - y$ values with sum mutations are normally distributed in the $x$, $y$ problem - simulation results.** Colored solid curves illustrate distributions of $x - y$ values at different time points. Dashed black curves show best fit (in terms of maximal likelihood) to Gaussian – with excellent agreement. Time $T$ is given in number of generations in the simulation. Simulation parameters: $\beta$ - selection with coefficient $\beta = 5$; mutations were normally distributed $N(0, 0.05)$. Population was initiated at the origin. Results based on 10,000 points.

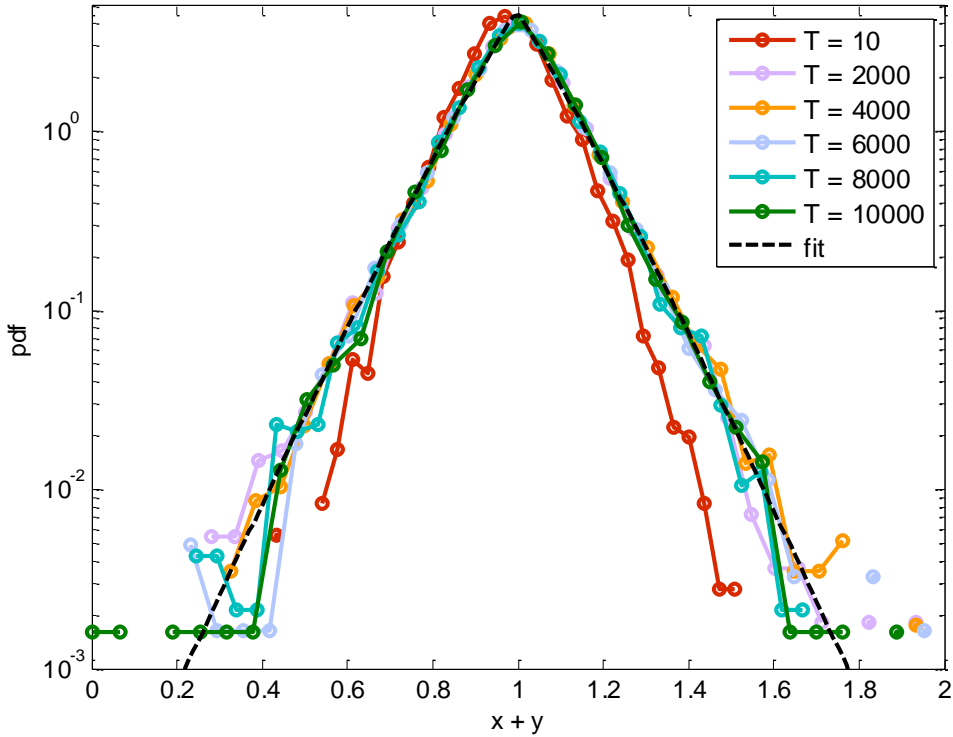

**Fig S3: $x + y$ distributions with sum mutations converge to a stretched exponential distribution - simulation results.** Colored solid curves illustrate distributions of $x + y$ values at different time points. Dashed black curve is fit to an effective potential $a \cdot e^{-b|x+y-1|^c}$ with a=4.6, b=11.1, c=1. All simulation points from $T = 5000$ were pooled together to produce the fit. Results pertain to the same simulations as in the previous figure.

### Dynamics under product-rule mutation near the sparse solution (0,1)

We now turn to product-rule mutations. By using the FP equation to describe mutations we assume that they are localized. This is a reasonable assumption for sum-mutations, but not for product-mutations, which can span a broad distribution of outcomes (see the "scaling effect" in the next section). To remedy this, we transform the equation to the logarithm of the original variables. Product is then transformed into sum, and the locality assumption of mutations is justified again.

If the product-rule mutation scheme is not symmetric, (in our case it is biased towards decreasing parameters values) a drift term should be added which is linear in the first derivative of the population density. Note that by the nature of the transformation to log-space, a product-rule that is biased towards decreasing values translates in log-space to a biased random walk towards $-\infty$. Using the transformation:

$$x = e^{G_x}$$

$$y = e^{G_y}$$

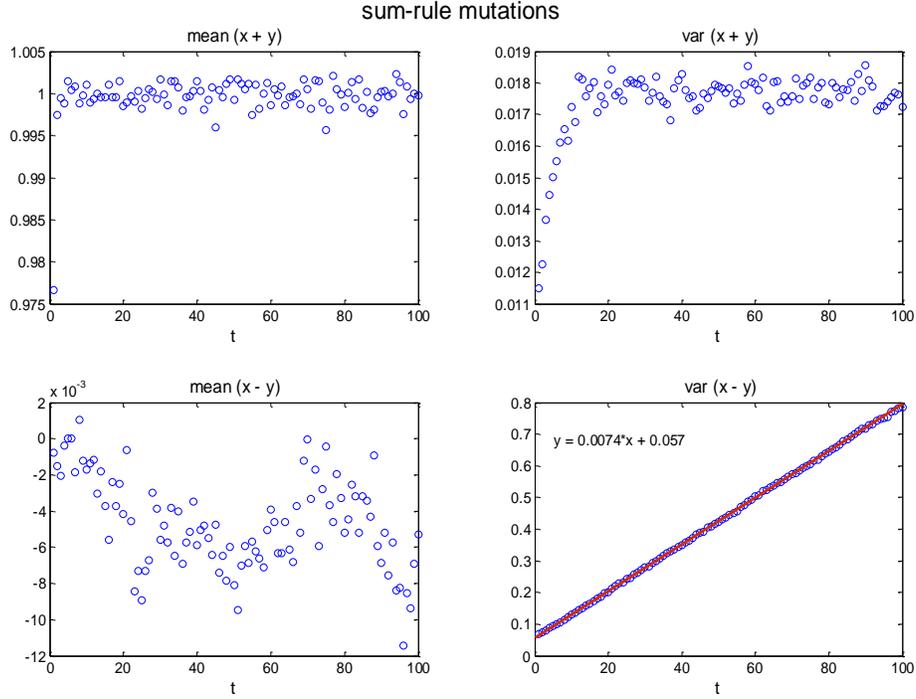

**Fig S4: Time dependence of the mean and variance of $x + y$ and $x - y$ in sum-rule simulations support our decomposition into two functional axes.** Selection rapidly drives the $x + y$ component to converge to the line $x + y = 1$ with constant variance. In contrast, the $x - y$ component freely diffuses, exhibiting a variance that grows linearly in time, in accordance with the analytical solution. The red line is a linear fit. Time $t$ is given in thousands of generations in the simulation.

where $G_x$ and $G_y$ are the log-transformed variables of x and y respectively. In equation:

$$\partial_t P(G_x, G_y, t) = D\, \nabla^2 P(G_x, G_y, t) - s \cdot \nabla P(G_x, G_y, t) - \nabla \cdot \left( P(G_x, G_y, t) \nabla F(G_x, G_y) \right).$$

Here $s < 0$ is the velocity of the drift towards $-\infty$. As with the symmetric case, this equation can be derived from the master equation under the assumption that the transition probability depends not only on the fitness difference between neighboring grid points, but also on the biased random walk probability to decrease. For example, in 1D the probability $p$ to move to the left (towards $-\infty$) is larger than ½. The value of $s$ in 1D is given by $(1 - 2p)\, \Delta G_x / \Delta t$. Note that in the symmetric case $= 1/2$, and $s$ naturally vanishes.

The fitness function $F$ (the potential in the FP equation) is:

$$F(G_x, G_y) = -\left(e^{G_x} + e^{G_y} - 1\right)^2.$$

We proceed by concentrating on one of the two sparse solutions to our problem (0,1). The sparse solution is obtained in the limit $(G_x, G_y) \to (-\infty, 0)$. The asymptotic form of the fitness in this limit again depends only on one of the variables

$$F(G_x, G_y) \to -(e^{G_y} - 1)^2.$$

With this in mind, the FP equation in log-transformed variables becomes:

$$\partial_t P = D\, \partial_{G_x} \partial_{G_x} P + D\, \partial_{G_y} \partial_{G_y} P - s\, \partial_{G_x} P - s\, \partial_{G_y} P - \partial_{G_y}(P\, \partial_{G_y} F(G_y)).$$

Using similar reasoning as in the sum-mutation case, we substitute $P = X(G_x, t)Y(G_y)$:

$$\partial_t X - D\, \partial_{G_x} \partial_{G_x} X + s\, \partial_{G_x} X = 0$$

$$D\, \partial_{G_y} \partial_{G_y} Y - \partial_{G_y}\left(Y \partial_{G_y} F(G_y)\right) - s\, \partial_{G_y} Y = 0$$

Similarly, the solution for the Y component is:

$$Y(G_y) = \frac{e^{(F(G_y)+s\, G_y)/D}}{\int e^{(F(G_y)+s\, G_y)/D}} \to \widetilde{Y}(y) = \frac{y^{-s/D} \cdot e^{\widetilde{F}(y)/D}}{\int y^{-s/D} \cdot e^{\widetilde{F}(y)/D}},$$

where the main difference is that the distribution is not necessarily symmetric. Here the population is concentrated around $e^{G_y} = y = 1$ with variance determined by $D$ and skewness determined by the ratio $-s/D$. Again note that in limit $s = 0$ the distribution is symmetric. The solution for the X component in log-space is again a Gaussian with variance that grows linearly in time, and a mean that moves to the left with velocity $s$:

$$X(G_x, t) = \frac{1}{\sqrt{4\pi D t}} e^{-\frac{(G_x + s\, t)^2}{4Dt}}.$$

Transforming to the original variables we found:

$$X(G_x, t)dG_x = \widetilde{X}(\log(x), t)\frac{dx}{x}$$

or

$$\widetilde{X}(t, x)dx = \frac{1}{x\sqrt{4\pi D t}} e^{-\frac{[\log(x)+s\, t]^2}{4Dt}} dx.$$

This is a lognormal distribution (5) with mode (most probable value) that converges to zero like $\exp((-s - 2D)t)$, but mean that diverges like $\exp((D - s)t)$, note that $s < 0$. For large t the leading term in the asymptotic expansion of this distribution goes like $\sim \frac{x^{\frac{-s}{2D}-1}}{\sqrt{4\pi D}} \frac{e^{-\frac{s^2}{4D}t}}{\sqrt{t}}$. This result with $s = 0$, agrees with the result in (6) that the product of infinitely many random variables converges to the log-normal distribution.

Simulation results demonstrating this behavior are shown in Figs. S5-S7.

In summary, we have shown by means of analytical solution and simulations that in the 2-variable toy model evolutionary dynamics with biased product-rule mutations bring us to solutions in which one of the variables asymptotically approaches zero. These are the sparsest solutions possible in this problem. These sparse solutions are strongly preferred although they show no fitness advantage relative to many other solutions that are equally fit but non-sparse. In contrast, with sum-mutations there is no preference to any specific solution as long as it achieves the goal. In the main text we show simulation results of a more complex matrix-multiplication model, which exhibits a very similar behavior. Under product mutations, the solutions obtained there are those that have the maximal number of zeros that still satisfy the goal, where under sum-mutations again arbitrary solutions that satisfy the goal are obtained. The likelihood of the latter solutions to be modular is very low.

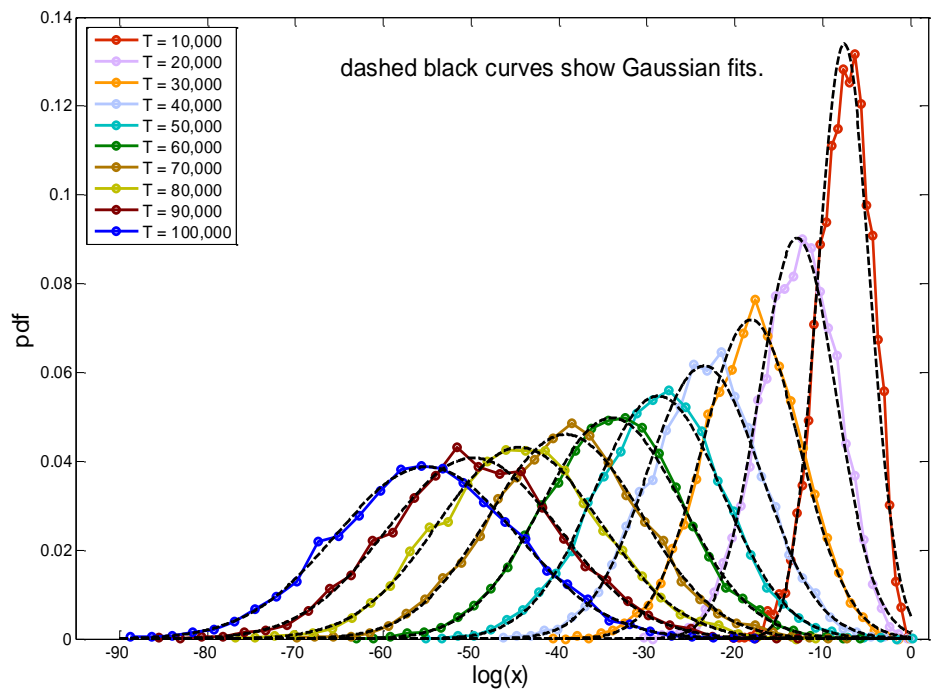

**Fig S5: x values under product mutations are log-normally distributed and asymptotically approach zero in the x, y problem - simulation results.** Colored solid curves illustrate distributions of log(x) values at different time points. Dashed black curves show best fit (in terms of maximum likelihood) to Gaussian – with excellent agreement. Time $T$ is given in number of generations. Simulation parameters: $\beta$-selection with $\beta = 5$, mutation normally distributed N(1,0.2). Population was initiated at the origin. In order to concentrate on one of the two sparse solutions, only simulation points with $0 < x < 0.5$ were considered in this analysis (roughly ~6000 points at each time point).

### Stochastic simulations of toy model

To estimate the temporal behavior of population distributions in this toy model we performed repeated runs of our simulation. At each run we randomly sampled a single individual from the population at the sampled time point. This was done in order to avoid dependence between different members of the same population due to finite population size. Each run was initiated with a different random seed, to assure independence of the distinct runs. Simulation consisted of repeated mutation-selection rounds, as described in the Methods Section of the main text. We used β-selection with $\beta = 5$. Mutations were normally distributed N(0,0.05) for sum rule mutations and N(1,0.2) for product rule mutations. Simulation was initiated with the population normally distributed around the origin $x = 0, y = 0$, with std 0.1 in both $x$ and $y$ axes.

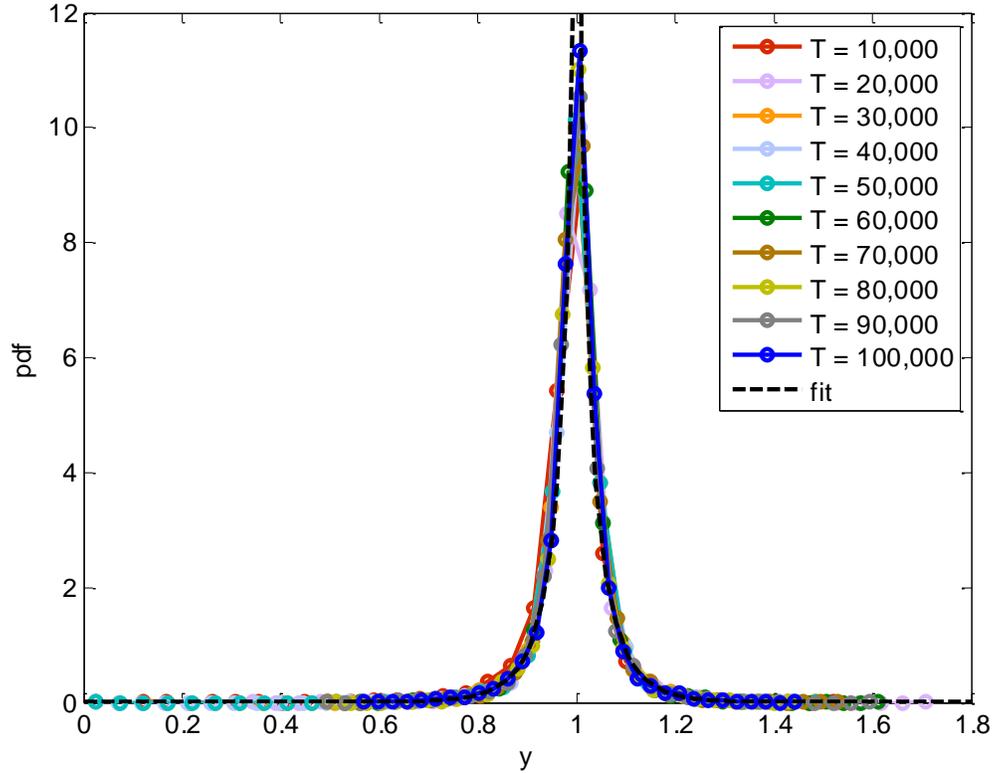

**Fig S6: y values with product mutations converge to a stretched exponential distribution in the $x, y$ problem - simulation results.** Colored solid curves illustrate distributions of y values at different time points. Dashed black curve shows fit to an effective potential $a \cdot e^{-b|y-1|^c}$ with a=21.19, b=14.8, c=0.67. Time T is number of generations. Results pertain to the same simulation points as in the previous figure (i.e. the y values corresponding to $0 < x < 0.5$).

The β-selection includes fitness scaling of the form: $f_i \to \frac{e^{-\beta(x_i+y_i-1)^2}}{\sum_i e^{-\beta(x_i+y_i-1)^2}}$. However, the relation to the potential is more complicated. Thus, we fit the simulation results to an effective potential of the form $b \cdot |x + y - 1|^c$.

Under these initial conditions the product-rule mutations have equal probability to converge to either one of the two sparse solutions. Simulation results are thus a superposition of the two solutions. When relevant to the analysis, we separated simulation points. To show the distribution approaching to the (0,1) solution we selected only points with x <0.5.

To plot distributions of product-mutation simulation we used uniform binning in the log domain. Fits to Gaussian are maximum likelihood estimators under the assumption that the data is normally distributed, calculated using the Matlab function 'normfit'.

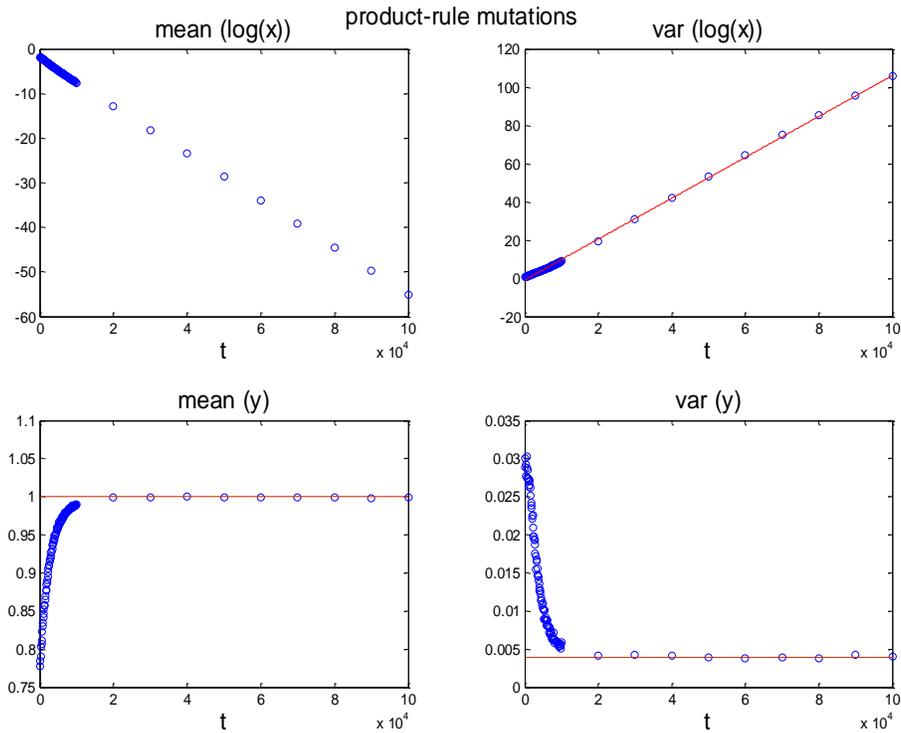

**Fig S7: Time dependence of x moments under product-rule mutations agrees with log-normal distribution predicted by analytical solution – simulation results.** The variance of log(x) is found to grow linearly in time and the mean of log(x) decreases linearly, as predicted by the analytical solution, assuming biased random walk in the log-space. Top row: red lines show best linear fits. In contrast, both the mean and the variance of y converge to a constant (red lines were added to guide the eye). Results pertain to the same simulations as in the previous two figures.

## 2. Mutation properties

### "Scaling effect" – sum vs. product mutations

Product mutations have the property that the pre-mutation value scales the distribution of potential outcomes. For example: multiplying the number 0.1 will result in a narrower distribution of potential outcomes compared to the one obtained if we multiplied the number 1 by values drawn from the same distribution. This property holds for both symmetric and asymmetric product mutations. Thus, the smaller is the pre-mutation value, the less likely it becomes to "escape" from it by mutation. Intuitively this explains why product mutations keep small interaction terms small.

Sum-mutations in contrast do not have this "scaling effect" - the distribution of mutation outcomes has the same width, regardless of the pre-mutation value. Thus product-mutations are fundamentally different from sum-mutations – see illustration in Fig. S8.

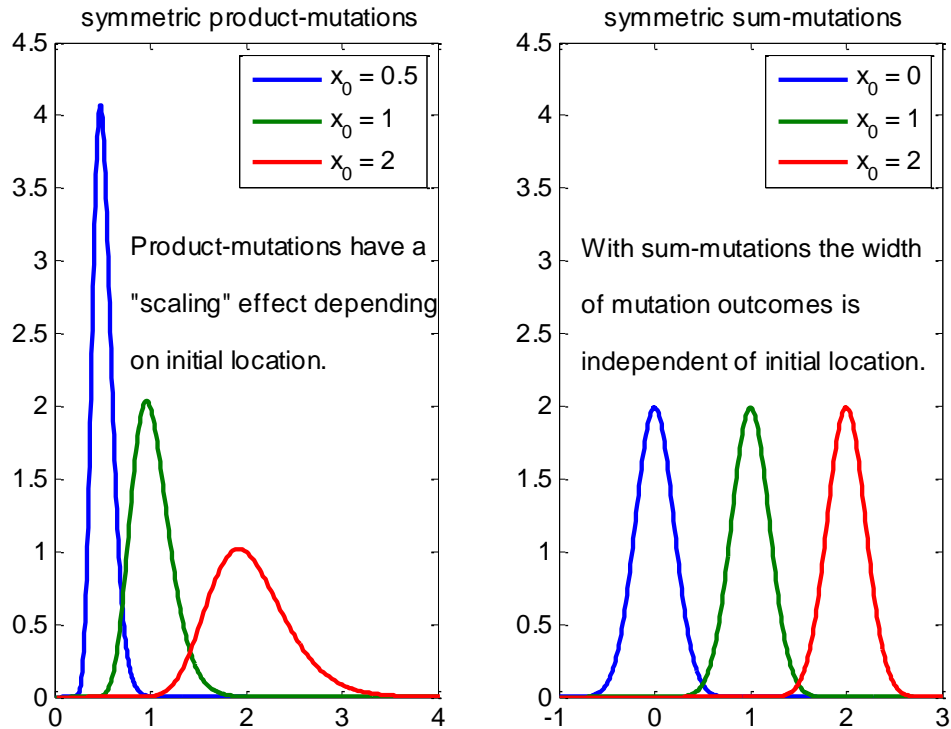

**Fig S8: Product-mutations have a scaling effect, but sum-mutations do not.** We compare the distribution of possible outcomes due to a single symmetric product-mutation (left) to that of a symmetric sum-mutation (right). We plot here the distribution of possible outcomes following such mutations to different pre-mutation values $x_0$. With sum-mutations the pre-mutation value has no effect on the width of the distribution of outcomes, and the distribution is simply relocated. In contrast, under product-mutations the smaller is the initial value, the narrower is the distribution of outcomes – that is the scaling effect. Thus the smaller are the values, the harder it is to escape. Here mutations are symmetric: either drawn from log-normal with $\mu=0$, $\sigma=0.2$ which have product symmetry (left) or drawn from normal distribution with $\mu=0$, $\sigma=0.2$ which has sum-symmetry.

## Symmetry of product-mutations

The (a)symmetry of mutations determines whether the center of the mutational distribution moves or not, which is a different effect. In biological mutations both the scaling and the relocation effects exist. The discussion of mutations symmetric with respect to product is thus purely theoretical. Several works have shown that mutations are biased to decrease interactions (see citations in the main text). Thus a realistic model should capture both effects: the product nature of mutations and their asymmetry. This discussion is meant to distinguish between the mathematical effects of these two properties, but not to argue that symmetric mutations are biologically relevant.

Which mutations are symmetric with respect to product?

To require symmetry with respect to product means that following many multiplications the geometric mean of the product will converge to 1:

$$\lim_{n\to\infty} \sqrt[n]{\prod_{i=1}^{n} x_i} = 1.$$

Taking the logarithm of this equation it is equivalent to:

$$\lim_{n\to\infty} \frac{1}{n}\sum_{i=1}^{n} \log x_i = 0.$$

Then by the law of large numbers $\log x_i$ is a random value with expectation zero. Assume that $\log x_i$ is normally distributed, then $x_i$ is log-normally distributed with parameter $\mu = 0$. To show this in an alternative way, assume that $x_i$ is distributed with probability density $f$ and equate the probabilities to multiply by $x$ and by $1/x$:

$$f_x(x) \cdot \Delta x = f_x\left(\frac{1}{x}\right) \cdot \left(\frac{1}{x} - \frac{1}{x+\Delta x}\right).$$

Assume that $\Delta x$ is small, we can approximate the interval $\left(\frac{1}{x} - \frac{1}{x+\Delta x}\right) \approx \frac{\Delta x}{x^2}$.

The equation becomes: $f_x(x) \cdot \Delta x = f_x\left(\frac{1}{x}\right) \cdot \frac{\Delta x}{x^2}$. Taking $f_x(x)$ to be the log-normal distribution, then:

$$\frac{e^{-(\log x - \mu)^2/2\sigma^2}}{x\sqrt{2\pi}\sigma} = \frac{xe^{-(\log(\frac{1}{x}) - \mu)^2/2\sigma^2}}{\sqrt{2\pi}\sigma} \cdot \frac{1}{x^2}$$

To satisfy this equation for every $x$ we obtain that $\mu = 0$, regardless of the value of $\sigma$.

Throughout this work, symmetric product-mutations were drawn from this distribution. The difference between the cumulative effect of symmetric and asymmetric product mutations is illustrated in Fig. S9.

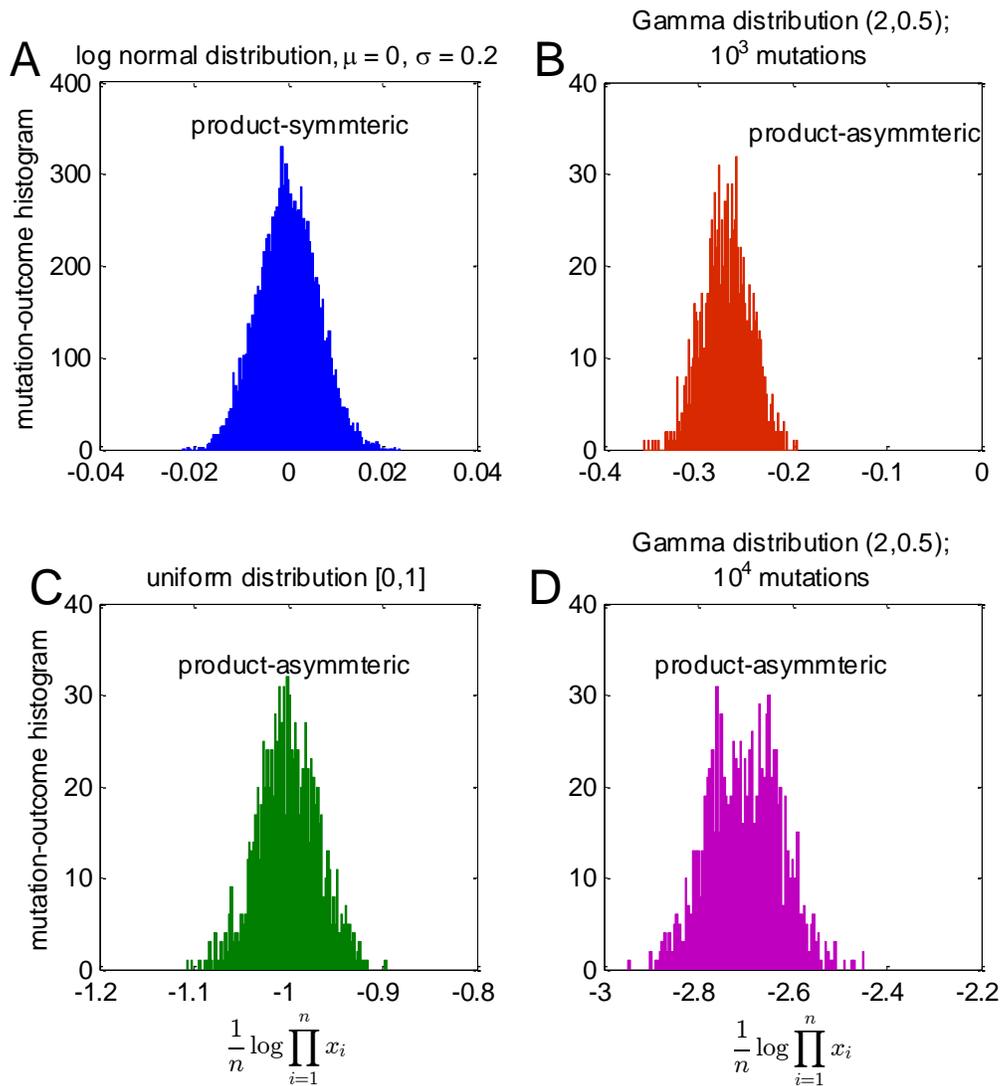

**Fig S9: We demonstrate the cumulative effect of symmetric and asymmetric product mutations in the absence of selection.** We assume that the initial value is 1, and multiply it by 1000 random numbers drawn from one of several distributions (detailed below). We plot here the histogram of the logarithm of cumulative mutation outcomes. (A) Product symmetric mutations were drawn from the log-normal distribution with **μ**=0, **σ**=0.2. Because mutations are symmetric with respect to product, the histogram is concentrated around 0 in the log-space (that is around 1 in the original variables), which indicates no mutational bias. (B) Mutations drawn from Gamma distribution are biased to decrease. Thus after 1000 multiplications the histogram is concentrated around a negative value in the log-space (value<1 in the original variables). (C) The uniform distribution [0,1] is also biased to decrease. (D) The bias increases with time. Here we multiplied by 10,000 random numbers drawn from the Gamma distribution (compare to B – with only 1000 multiplications). The illustrated histograms are based on 1000 points each.

**Both symmetric and asymmetric mutations lead to sparse/modular solutions:**

The symmetric and asymmetric mutations differ in their effect if we had only mutations active, but not selection. In the absence of selection asymmetric mutations will bring all interactions to near-zero (given enough time), but symmetric mutations will not. If selection is active too, both symmetric and asymmetric mutations will result in sparse/modular solutions. With symmetric mutations, the reason for this is that selection breaks the symmetry. Under product-rule, mutations in the finite interval [0,1] are compensated by mutations in the infinite interval $[1, \infty)$. Thus selection for some finite goal value will always create a bias towards lower values, and thus produce a tendency to decrease, similarly to asymmetric mutations. In addition, our simulations have demonstrated that the "symmetric" state is also unstable: even a slight asymmetry is sufficient, because of the enormous number of generations in our simulations.

We illustrate below simulation results with product mutations, both symmetric and asymmetric in the $x + y = 1$ problem, described in the previous section. As can be seen, both mutation types lead to sparse solutions, but with asymmetric mutations this effect is naturally stronger (Figs. S10-S11).

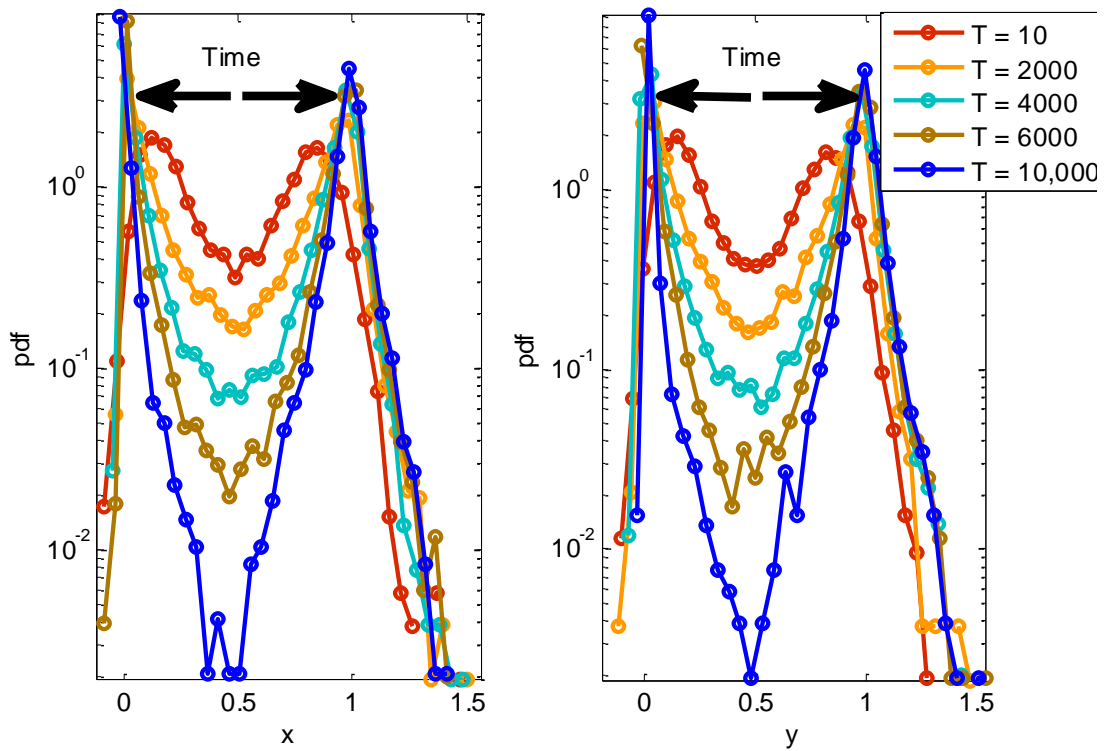

**Fig S10: Evolution with asymmetric mutations in the $x + y = 1$ problem.** Mutations were drawn from a normal distribution N(1,0.2). The population distributions of $x$ and $y$ values at several time points are illustrated. As time goes on, the distributions become more and more concentrated around the sparse solutions (0,1) and (1,0). Simulation conditions are the same as in the previous section.

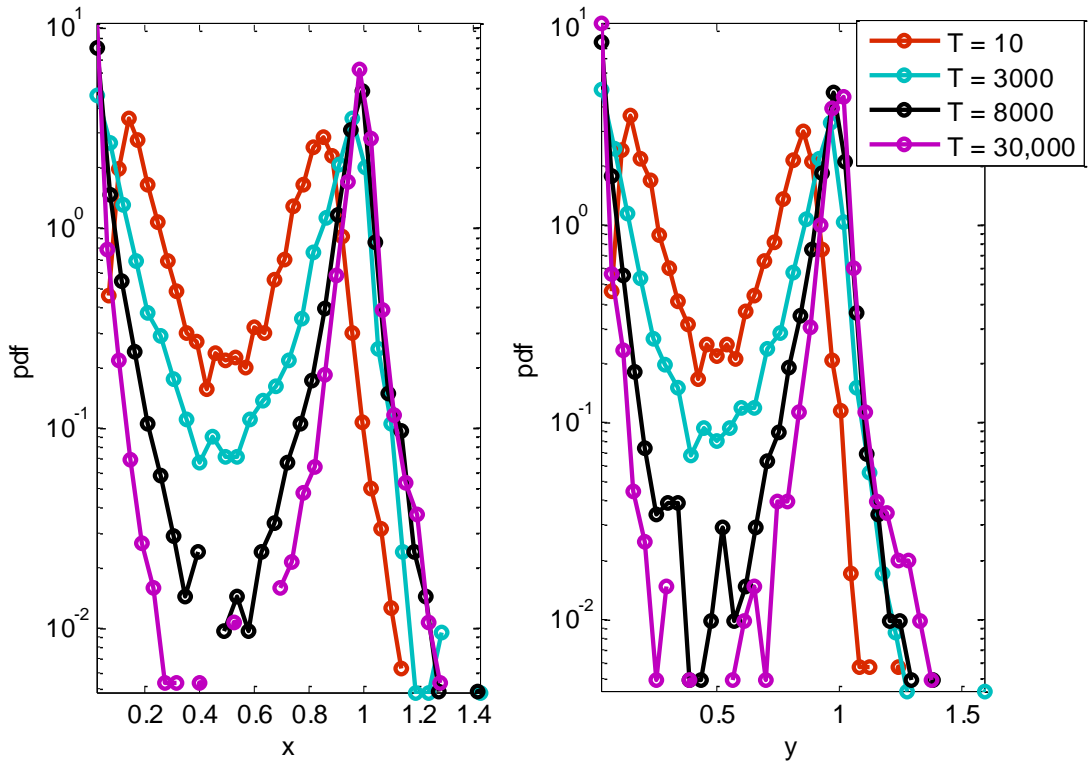

**Fig S11: Evolution with product-symmetric mutations in the $x + y = 1$ problem.** Mutations were drawn from a log-normal distribution LN(0,0.2). The population distributions of $x$ and $y$ values at several time points are illustrated. As time goes on, the distributions become more and more concentrated around the sparse solutions (0,1) and (1,0). Simulation conditions are the same as in the previous section.

### 3. Evolutionary simulations

Here we detail our evolutionary simulations. Simulations were written in Matlab using a standard framework (7, 8). We initialized the population of matrix pairs by drawing their $N \cdot 2D^2$ terms from a uniform distribution at either small range, i.e. $\mathcal{U}[0, 0.1]$ or large range $\mathcal{U}[0,1]$- where small and large are compared to the largest elements in the goal matrix (which are of order one). In other simulations we used initial values drawn from a normal distribution around zero with std 0.1 or 1. Most results shown refer to the small range, which relates to evolution of a structure with only weak initial interactions; however our conclusions apply also to the large range. Population size was set to $N = 500$.

At each generation the population was duplicated. One of the copies was kept unchanged, and elements of the other copy had a probability $p$ to be mutated – as we explain below. We note that it is also common to keep a single copy of the population and mutate it. The former technique is less likely to lose a good solution once it is found, but its convergence is relatively slow, whereas the latter technique is faster, but might lose beneficial solutions that have already been found (see discussion in (7) chap. 10). Fitness of all $2N$ individuals was evaluated by $F = -||AB - G||$, where $||\cdot||$ denotes the sum of squares of elements (Frobenius norm). This formula represents the Euclid distance of the matrix product from the goal (9). The best possible fitness here is zero, achieved if $AB = G$ exactly. Otherwise, fitness values are negative. In the figures we show the absolute value of mean population fitness. The goal matrix was either diagonal $G = 2 \times I$, nearly-diagonal (diagonal matrix with small non-diagonal terms), block-diagonal or full rank with no zero elements. $N$ individuals were then selected out of the $2N$ population of original and mutated ones, based on their fitness. This mutation–selection process was repeated again and again until the simulation stopping condition was satisfied (usually when mean population fitness was less than 0.01 from the optimum).

**Mutation**: We tested point mutations in our simulation and assumed statistical independence between mutations at different elements. We kept mutation rate such that on average 10% of the population members were mutated at each generation, so the element-wise mutation rate $p$ for matrices of dimension $D$ was at most $\frac{0.1}{2D^2}$. This relatively low mutation rate enables beneficial mutants to reproduce on average at least 10 times before they are mutated again. In simulations where we compared dependence on matrix dimension (Fig. 5) we used the same mutation rate at all dimensions, generally the one that pertains to the highest dimension used in the simulation.

We randomly picked the matrix elements (in both $A$ and $B$) that would be mutated. Mutation values were drawn from a Gaussian (or log-normal or Gamma) distribution. For sum-rule mutation, this random number was added to the mutated matrix value: $A_{ij} \to A_{ij} + \mathcal{N}(0,\sigma)$ or $B_{ij} \to B_{ij} + \mathcal{N}(0,\sigma)$, and for product- rule, the mutated matrix element was multiplied by the random number: $A_{ij} \to A_{ij} \cdot \mathcal{N}(\mu,\sigma)$ or $B_{ij} \to B_{ij} \cdot \mathcal{N}(\mu,\sigma)$. Mean mutation value $\mu$ was usually taken as 1, however we also tested other values of $\mu$ (both larger and smaller than 1) and results remained qualitatively similar, only the time-scales changed.

When we tested mutations which are symmetric with respect to product we took the log-normal distribution with $\mu = 0$.

We also tested the dependence on the mutation size $\sigma$, using $\sigma = 0.01 - 3$, and found similar results. In most simulation results shown here we used $\sigma = 0.1$ (unless stated otherwise). Fitness convergence crucially depends on the mutation frequency and size, as demonstrated in our sensitivity test. Grossly speaking, a high mutation rate can speed up evolution at the beginning of the simulation, but can later on preclude slightly better mutants from taking over, because they are mutated again before they reproduce sufficiently. There is also a similar trade-off with mutation size: large mutations can speed evolution at the beginning, but at the final stages the mutation size limits the precision with which the goal can be approached.

**Selection methods**: We tested 3 different selection methods; all gave qualitatively very similar results with only difference in time scales. Most results presented here were obtained with tournament selection (see (7) chap. 9): $N$ sets, each containing $s$ population members, were uniformly drawn with repetitions. The best individual at each set was then selected to be at the population next generation. This mimics the fact that an individual needs to outperform only others at its close vicinity, rather than the whole population. The parameter $s$ can be used to tune the selection intensity (the larger it is, the stronger is the selection). In our simulations we set $s = 4$.

Another selection method tested is "truncation-selection" or "elitism". Here population members were ranked by their fitness. The best half of members were selected and duplicated. We note that both methods are based on the fitness rank, rather than on its exact value, making fitness scaling unnecessary. Both methods gave very similar results.

The third method used was proportionate reproduction with Boltzmann-like scaling (10–12): here the relative fitness was computed as $\tilde{F}_i = e^{\beta F_i}/\sum_j e^{\beta F_j}$. Evidently $\sum_j \tilde{F}_i = 1$, so that $\tilde{F}_i$ is the probability of the $i$-th individual to be selected. The parameter $\beta$ determines the selection strength, where at one extreme if $\beta = 0$, all individuals are equally probable to be selected and at the other extreme if $\beta = \infty$, the best individual is selected with probability 1, while all others have probability zero to be selected. To implement selection we then exploited the "roulette-wheel" algorithm (7, 8) where a section of the interval [0,1] equal to $\tilde{F}_i$ was assigned to the $i$-th individual. $N$ Random numbers were then uniformly drawn from the interval [0,1]. The individuals whose sections contained such numbers were then selected (with repetitions).

For a comparative test of the dependence of fitness achieved and the time needed to reach it on selection and mutation parameters see sensitivity test below.

If selection is too weak (e.g. $\beta = 0.1$ in Boltzmann-like selection) sparse structures are obtained, but their fitness is far from optimal. If the fraction of individuals mutated at each generation is too high (e.g. every individual has on average one mutation per generation), then again the solutions obtained are bounded away from the optimum, because high fitness individuals are likely to suffer from additional deleterious mutation before they reproduce sufficiently.

## 4. Evolutionary simulation parameter sensitivity test

Here we show in Fig. S12 the dependence on mutation size and selection intensity β of the evolutionary simulation with the Boltzmann-like selection scheme. In this test we let the simulation solve a 1-D problem for a fixed number of generations (=800), with a single repeat for each parameter combination. We tested 6 different values of β (0.1-20) and 5 different values of the mutation size σ (0.01-0.5). Here we plot either the mean population fitness (top row: **A** and **B**) or the best fitness obtained within the population (bottom row: **C** and **D**), reached within this fixed number of generations. In the left panels (**A** and **C**) each curve illustrates the dependence on β for a fixed mutation size, and the right panels (**B** and **D**) show the dependence on mutation size where each curve was obtained for a different values of β. Curves in both left and right panels were created by the same simulation results.

Alternatively, we tested how the time to reach a desired fitness (0.01 from the optimum) depends on these parameters in a 3-$D$ problem. The number of generations was limited to 500,000 and some parameter combinations failed to reach the required fitness by that time. Similarly, we show in Fig. S13 the dependence on $\beta$ for fixed mutation size (**A**) or dependence on mutation size for fixed $\beta$ (**B**).

Based on these tests we chose to set the mutation size $\sigma = 0.1$ and the selection intensity $\beta = 10$ (Boltzmann-like) or $s = 4$ (tournament).

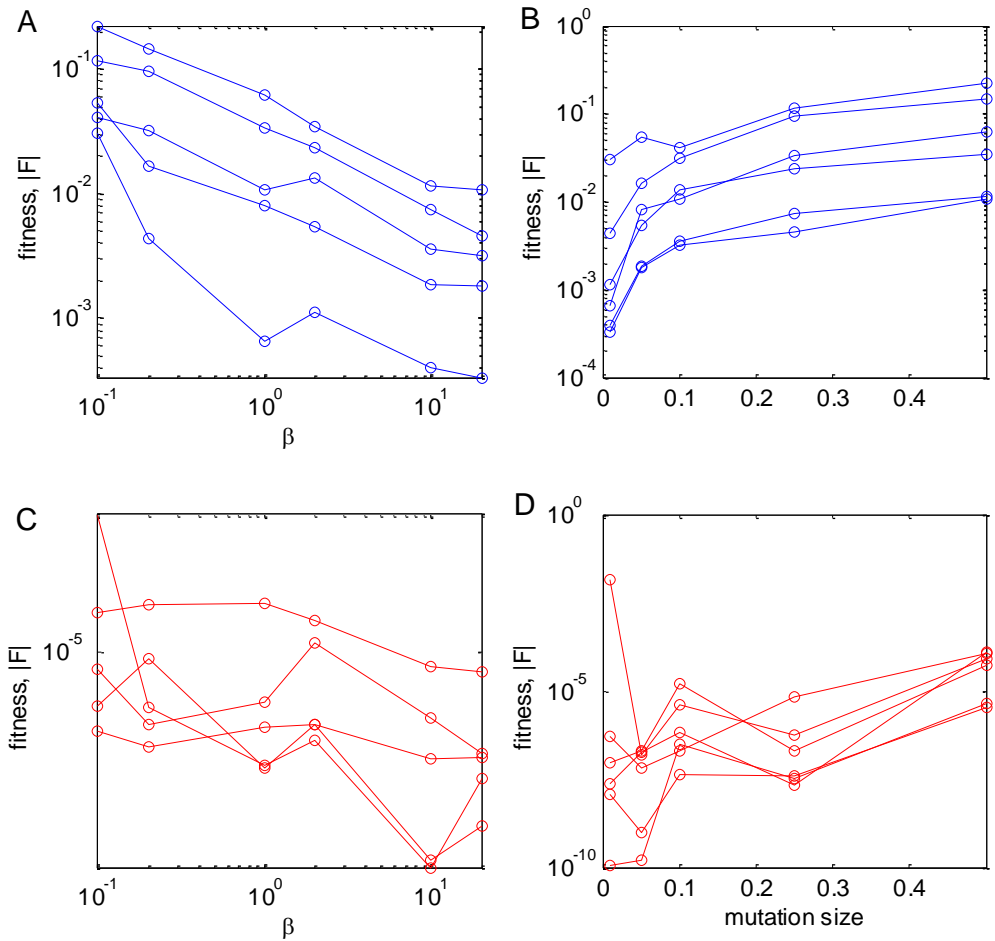

**Fig S12: Dependence of the achieved fitness on selection strength and mutation size in a 1-D problem.**

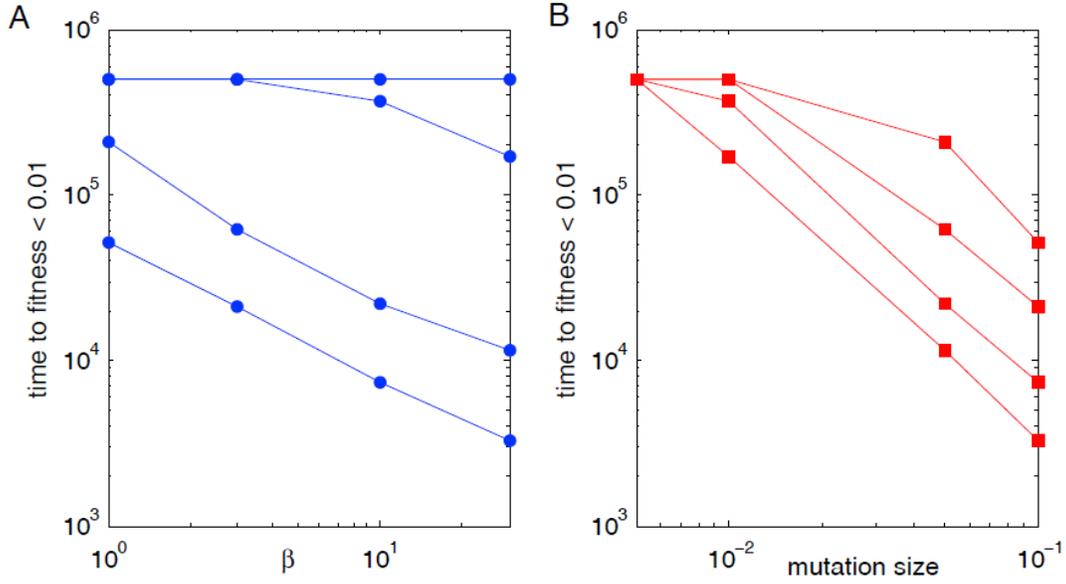

**Fig S13:** dependence of the time to reach a desired fitness value on selection strength and mutation size.

## 5. Modularity: definitions and error calculation

**Definition of modularity**: if the goal is diagonal, we define modularity as $M = 1 - \langle|n|\rangle/\langle|d|\rangle$ where $\langle|n|\rangle$ and $\langle|d|\rangle$ are the mean absolute value of the non-diagonal and diagonal terms respectively. At each generation, the $D$ largest elements of each matrix (both $A$ and $B$), were considered as the diagonal $\langle|d|\rangle$ and the rest $D^2 - D$ terms as the non-diagonal ones $\langle|n|\rangle$. Averages were taken over matrix elements and over the population. This technique copes with the unknown location of the dominant terms in the matrices, which could form any permutation of a diagonal matrix. Thus, $0 \leq M \leq 1$: where at the two extremes, a diagonal matrix has $M = 1$, and a matrix with equal terms has $M = 0$. Since we choose the largest elements to form the diagonal, negative values of $M$ are not allowed. When the goal is non-diagonal, one can use standard measures for modularity such as (13) [not used in the present study].

**Calculation of time to modularity**: we used the following approximation for fitness value when the goal is diagonal. Assume that $A$ and $B$ are $D$-dimensional matrices consisting of 2 types of terms: diagonal terms all with size $d$ and non-diagonal terms all with size $n$ and that the goal is $G = g \times I_{D \times D}$. The fitness then equals:

$$-F = D[d^2 + (D-1)n^2 - g]^2 + D(D-1)(2dn + (D-2)n^2)^2.$$

We collect terms by powers of $n$, and obtain a constant term and terms with powers $n^{2,3,4}$. Modularity is obtained when the solution has the correct number of dominant terms appropriately located and their size is approximately $d^2 \cong g$. At the beginning of the temporal trajectory, when non-diagonal elements are relatively large, $F$ is dominated by the $O(n^4)$ term. When a modular structure emerges, non-diagonal elements become relatively small, and the dominant term remaining in $F$ is $O(n^2)$. Our criterion for determining time to modularity was the time when the $n^2$ term first became dominant, i.e. when $F - n^2(\cdots) < n^2(\cdots)$.

**Matrix permutations:** For ease of presentation we permuted the $A$ and $B$ matrices, so that they form nearly-diagonal matrices. This applies to the cases when $G$ is diagonal and $A$ and $B$ also evolve to be (nearly) diagonal. We used the same permutation for the rows of $A$ and columns of $B$. Such permutation preserves the matrix product and is equivalent to simply changing the order of inputs. To find the correct permutation, we sorted each column of $A$ in descending order. Then the first row in the sorted matrix had the $D$ largest elements. We used the order vector of this first row (i.e. indices of rows where these elements were located in the original $A$) as the required permutation.

**Calculation of error bars in time dependence on $D$:** We repeated the simulation at each dimension either $K = 140$ times ($D = 3 - 10$) or $K = 80$ times ($D = 11 - 15$), initializing the Matlab random seed with a different integer number each time. At each run we measured time to reach fitness within 0.01 of the optimum and time to modularity, as explained above. As these times formed a broad and highly skewed distribution, we considered their median, rather than their mean. To estimate our error in this median estimator, we used the following bootstrapping procedure. We randomly formed sets of $K$ samples (with repetitions) of simulation results. We constructed $L = 10{,}000$ such sets, and calculated the median of each. We then calculated the standard deviation of these median values. To estimate the error in the dependence of the time on $D$, we randomly picked one measurement from each dimension and then calculated the best line (in terms of least squares) connecting these points. We repeated this process 10,000 times, receiving each time different parameters for the best line. Errors in line estimation presented here, represent the 5% and 95% quantiles out of the obtained distribution of line parameters.

## 6. LU decomposition - proofs

An LU decomposition exists for every full rank matrix (14). In such decomposition there is a total of $D^2 - D$ zeros in both A and B together. Here we prove that a larger number of zeros is not possible unless G has a zero term (or is not full rank).

The $D^2 - D$ zeros can be partitioned between A and B in different ways: either equally (the LU decomposition, where A and B are triangular matrices), or all zeros in one of the matrices and none in the other or any other partition.

**Theorem: maximal number of zeros in LU decomposition of a full rank matrix with no zero elements is $D^2 - D$.**

**Lemma**: Let $A_1 B_1 = G = A_2 B_2$ be 2 different decompositions of the goal $G$ with different zero partitions, such that all matrices are invertible. Then, there exists an invertible transformation matrix $P$, such that $A_1 P = A_2$ and $P^{-1} B_1 = B_2$.

**Proof**: Define $P = (A_1)^{-1} A_2$. Then $A_1 P = A_1 (A_1)^{-1} A_2 = A_2$ and $P^{-1} B_1 = (A_2)^{-1} A_1 B_1 = (A_2)^{-1} G = (A_2)^{-1} A_2 B_2 = B_2$.

Q.E.D

If a transformation exists between all pairs of decompositions, specifically we can choose $A_2 B_2$ in which $A$ is full and $B$ is diagonal, i.e. all $D^2 - D$ zeros are in $B$. Now let's check what happens if we try to add one more zero. Then, because $B$ is diagonal, $G_{ij} = \sum_k A_{ik} B_{kj} = A_{ij} B_{jj}$, $\forall i, j$. Without loss of generality we set $A_{ij} = 0$, then essentially $G_{ij} = 0$, so $G$ is not a general matrix.

Alternatively if we set $B_{jj} = 0$, we will obtain that the $j$-th column of $G$ is all zeros – hence $G$ is not full rank.

Q.E.D

**Theorem: If $G$ is full rank but has $k$ zeros, the maximal number of zeros in LU decomposition is $D^2 - D + k$.**

As stated above, for a general full rank $G$ a decomposition in which $A$ is full and $B$ is diagonal, (i.e. there is a total of $D^2 - D$ zeros) is possible.

Now assume without loss of generality that $G_{ij} = 0$. Since $B$ is diagonal $G_{ij} = A_{ij} B_{jj}$, so that $A_{ij}$ must be zero too ($B_{jj} \neq 0$ because otherwise a full column in $G$ equals zero and then $G$ is not full rank). Consequently, for every zero in $G$, we obtain exactly one additional zero in $A$, which proves our claim that for $G$ with $k$ zeros, we obtain a decomposition with exactly $D^2 - D + k$ zeros.

Due to the lemma above, these zeros can be split in different ways between $A$ and $B$.

Q.E.D

## 7. Nearly modular $G$ - supplementary figure

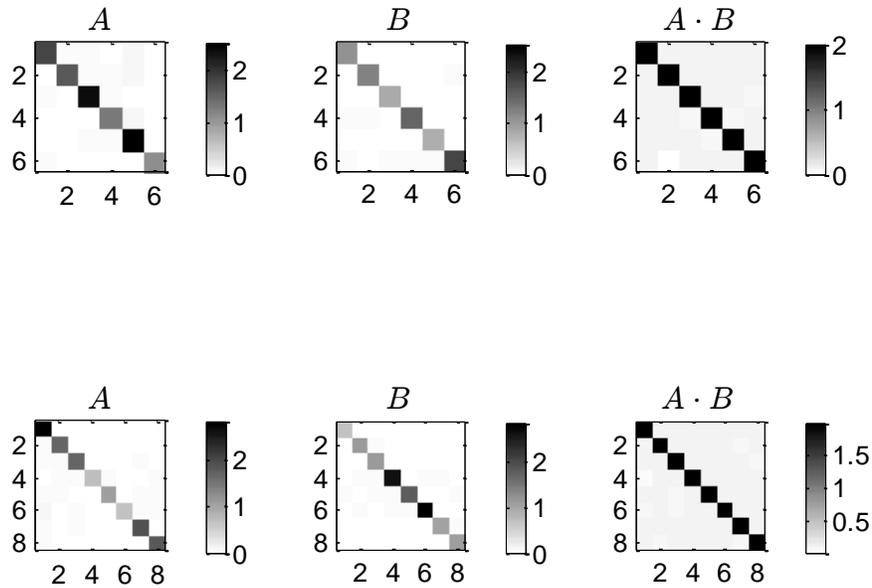

**Fig S14: If the goal $G$ is nearly diagonal, the evolutionary simulation with product-rule mutations reaches solutions in which $A$ and $B$ are nearly-diagonal too.** We set $G$ to be a matrix with values of 2 on its diagonal and 0.1 in its all non-diagonal terms. Here we show two examples of solutions obtained for $D = 6$ (top row) and $D = 8$ (bottom row). Numerical values are represented by color code when white represents zero. Matrices were permuted to form the most diagonal form (see above).

## 8. Mutation sign and distribution – supplementary figure

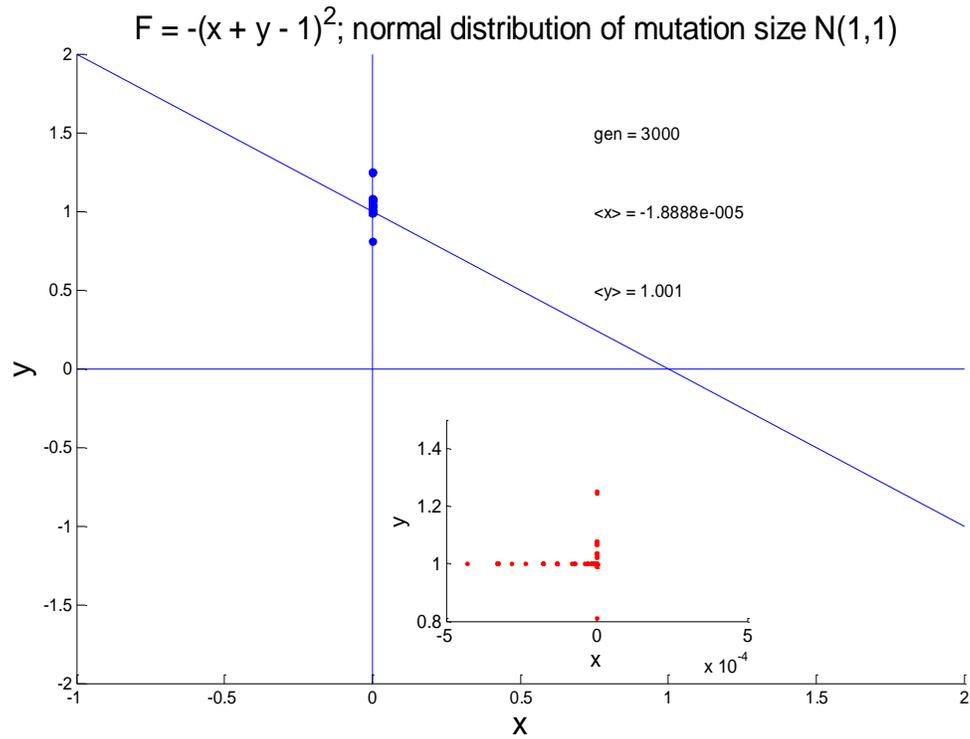

**Fig S15: Broad distribution of mutation values allows for negative as well as positive matrix values.** Here we show the distribution of solutions to the $x, y$ problem with product mutations normally distributed N(1,1). The solutions concentrate near the modular point (0,1). Inset demonstrates that the x values are in fact negative in this case. Simulation was run for 3000 generations. Mean x and y values are written on top of the graph.

## 9. Block diagonal goal – supplementary figure

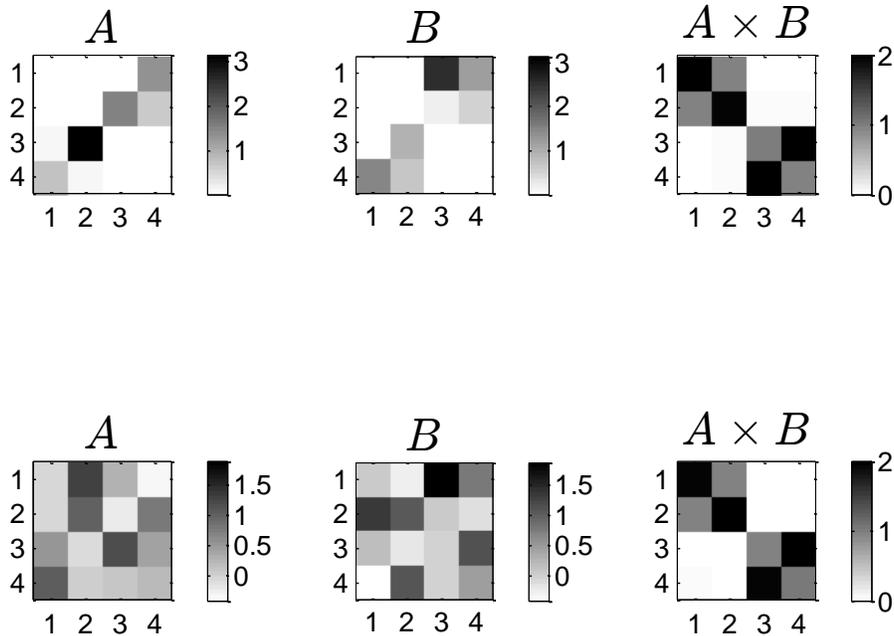

**Fig S16: Comparison of product-rule vs. sum-rule mutations over a block-diagonal goal.** The goal matrix here was the same block-diagonal goal as in Fig 2C (main text). Here we compare the different solutions obtained with product mutations (top row) to those obtained under sum-mutations (bottom row) with such a goal. Under product mutations, each block of the goal matrix is decomposed into a product of two triangular matrices – as happens for a general goal matrix. Under sum-mutations, we obtain non-modular solutions, as we did for diagonal goal matrices (compare to Fig.3 – main text).

**Supplementary movies**

Movies demonstrate the simulation dynamics in the $x, y$ problem under product-rule mutations with various distributions of mutations. All distributions converge to either of the sparse solutions.

**Movie S1**: Mutations had Gamma distribution with parameters Gamma(1, 40.25). In addition, each mutation value was multiplied by -1 with probability 0.1, so that matrix values could also change their sign. $\beta$- selection was used with $\beta = 5$.

**Movie S2**: Mutations had log-normal distribution with parameters LN(-0.11, 0.47). In addition, each mutation value was multiplied by -1 with probability 0.1, so that matrix values could also change their sign. $\beta$- selection was used with $\beta = 5$.